\documentclass[12pt]{elsarticle}

\usepackage{papersettings}
\usepackage{mathsettings}
\usetikzlibrary{fit}
\usetikzlibrary{angles}
\usepackage{setspace}

\newcommand{\fft}[1]{\widehat{#1}(\boldsymbol{ \xi})}
\newcommand{\ifft}[1]{\mathcal{F}^{-1}\{#1\}}

\graphicspath{{./figures/}}

\journal{}

\newcommand{\myreferences}{./references}
\newcommand{\gradient}{\vector{\nabla}}

\begin{document}

\begin{frontmatter}
  \title{An FFT-based framework for predicting corrosion-driven damage
  in fractal porous media}
  
    \address[ifbaddress]{Institute for Building Materials, ETH Zurich, Switzerland}
  \author[ifbaddress]{Mohit Pundir}
  \ead{mpundir@ethz.ch} 
  \author[ifbaddress]{David S. Kammer \corref{cor1}} \ead{dkammer@ethz.ch}
  \author[ifbaddress]{Ueli Angst}
  \cortext[cor1]{Corresponding author}

\begin{abstract}
  Understanding fracture in cementitious materials caused by the deposition and growth of corrosion products requires scale-bridging approaches due to the large length-scale difference between the micro-pores, where deposition occurs, and the structure, where deterioration manifests. Cementitious materials bear a highly heterogeneous micro-structure owing to the fractal nature of micro-pores. Simultaneously, a corrosion-driven fracture is a multi-physics problem involving ionic diffusion, chemical reactions, and stress development. This multi-scale and multi-physical character makes scale-bridging studies computationally costly, often leading to the use of simplified fractal porous media, which has important consequences for the quantitative interpretation of the results. Recent advances in homogenization approaches using Fast-Fourier-Transform (FFT) based methods have raised interest due to their ease of implementation and low computational cost. This paper presents an FFT-based framework for solving corrosion-driven fractures within fractal porous media. We demonstrate the effectiveness of the Fourier-based spectral method in resolving the multiple corrosion-driven mechanisms such as ionic diffusion, stress development, and damage within a fractal porous microstructure. Based on the presented methodology, we analyze the impact of simplifying fractal porous media with simple Euclidean geometry on corrosion-driven fracture. Our results demonstrate the importance of preserving both the porosity and fractal nature of pores for precise and reliable modeling of corrosion-driven failure mechanisms. 
\end{abstract}


\end{frontmatter}

\newpage


\section{Introduction}

\begin{figure}
  \centering
\begin{tikzpicture}
  \node[inner sep=0pt] at (0, 0) {\includegraphics[trim=0cm 0cm 0cm 0cm, clip, width=\linewidth]{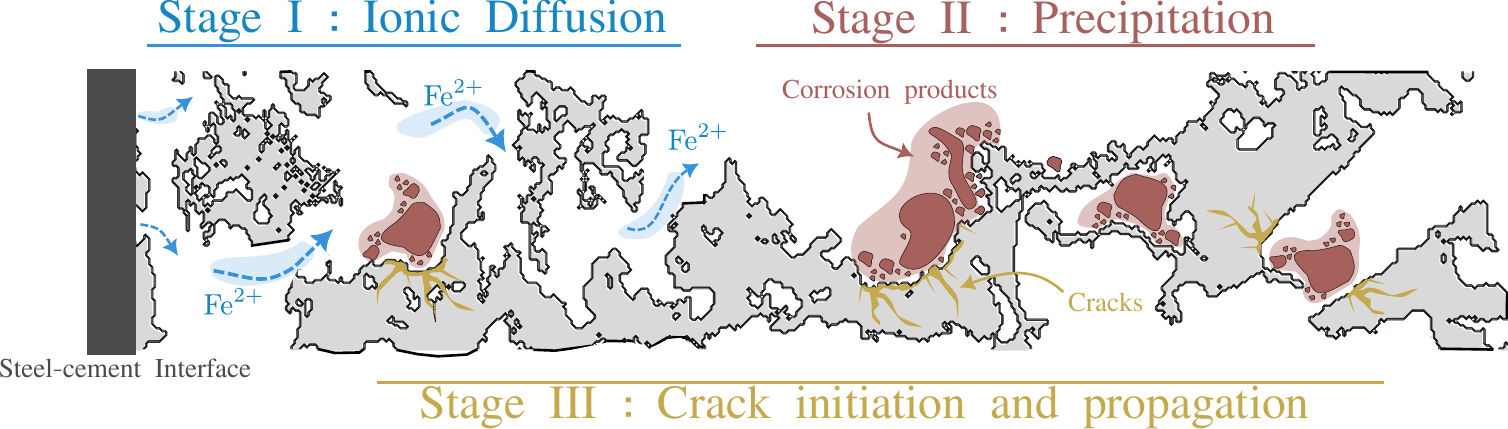}};
\end{tikzpicture}
\caption{\textbf{Corrosion-driven mechanisms.} Schematic figure showing various corrosion-driven mechanisms that lead to fracture in porous media. The entire corrosion process is divided into 3 stages. \textbf{Stage I} : The ferrous ions released at the steel-concrete interface undergo diffusion through the pore space (shown in white). \textbf{Stage II} : Over time, the ferrous ions precipitate and the precipitates grow within these pores. \textbf{Stage III} : The growth of the precipitates results in the development of stresses within the solid (shown in grey) and, consequently, crack initiation and propagation.}
\label{fig:conceptual}
\end{figure}

Corrosion of steel reinforcement bars in concrete plays a significant role in a structure's durability and serviceability lifetime~\cite{angst_challenges_2018, noauthor_corrosion_2003}. Ferrous ions released at the steel-concrete interface diffuse through the pores in the concrete and undergo many chemical reactions leading to the precipitation of corrosion products, as schematically illustrated in \Cref{fig:conceptual}. These precipitates, which are confined in the pore space, grow over time, exerting pressure onto the pore walls, subsequently leading to internal cracking around the pore space and eventually to macroscopic cracks. Even though the underlying processes are generally well recognized for their contributions to the degradation of steel-reinforced concrete, a precise and quantitative description of this multi-physical process remains missing. One major reason is that the pore space in concrete is highly complex~\cite{ulm_is_2003, ioannidou_mesoscale_2016}. The pore sizes range from the nanometer-scale to the micrometer-scale, and pore surfaces are fractal, \ie{} features exhibit similar patterns across all of these length scales. The complex and concealed nature of pore spaces thus makes it difficult to assess the aforementioned corrosion-driven mechanisms and the induced damage from external examination of structures. Therefore, numerical simulations have been an important tool to study the corrosion-driven mechanisms at the pore scale and to analyze their effect on a structure's durability and its serviceability lifetime.

These approaches often rely on synthetic representations of the pore space built from data derived from measurement techniques such as Mercury Intrusion Porosimetry (MIP), nitrogen-adsorption~\cite{zhu_fractal_2019,gao_fractal_2014,pia_geometrical_2013,zeng_surface_2013,yang_fractal_2017,jilesen_three-dimensional_2012,zhang_determination_1995,winslow_fractal_1995} or tomography ($\mu-$CT, FIB-SEM)~\cite{okabe_pore_2005, wang_study_2009}. These synthetic pore spaces are then employed in numerical frameworks, such as the finite-element method, pore network model, and Lattice Boltzmann method, to simulate various physical mechanisms within representative volumes. To bridge the physical mechanisms within the micro-pores to a structure's degradation at the macroscale, a multi-scale homogenization approach is followed. In these frameworks, the pore space is modelled either implicitly~\cite{noauthor_microporomechanics_2006} or explicitly~\cite{geers_multi-scale_2010}, where 
the complex pore space is often simplified using simple geometries such as spheres, cylinders and ellipsoids~\cite{liu_multi-scale_2020, maekawa_multi-scale_2014}. 
The main purpose is to reduce the computational complexities (high computational cost, complex analytical formulations) associated with high-resolution simulations of diffusion, stress development and crack initiation in a fractal domain. Although this idealization of a fractal space into a smooth Euclidean space (cylinder/sphere) is necessary, it approximates the studied physical process. For example, idealizing a fractal surface with a smooth surface approximates the stress development and the effective diffusion path of ferrous ions in the porous material, which affects the precipitation of corrosion products and the initiation of cracks. Therefore, an accurate representation of the pore space is necessary to fully capture and understand the influence of corrosion-driven mechanisms at the structural scale.  In this paper, we will answer the question if it is possible to preserve the actual representation of the pore space and simulate the corrosion-driven mechanisms in a computationally efficient manner. Thus, the aim is to present a single numerical framework for corrosion-driven processes that simultaneously includes the diffusion of chemical species, the stress development due to the growth of precipitates, and crack initiation, while being computationally less demanding than conventional approaches and being straightforward to implement for fractal spaces. Such a single framework allows for studying the interplay among all the mechanisms involved in this multi-physics-driven fracture process.  

The paper is organized as follows: In \Cref{sec:method}, we present the numerical framework based on the spectral method~\cite{moulinec_numerical_1998, de_geus_finite_2017, leute_elimination_2022, Bonnet2017} to simulate multiple corrosion-driven mechanisms simultaneously in a given microstructure: diffusion within the pore space, the stress development due to the growth of precipitates and the crack initiation in the matrix.  In \Cref{sec:results}, we employ the presented framework to simulate the mechanisms above in fractal pore spaces reconstructed from tomographic scans of cementitious materials. We apply the proposed framework to a multi-scale setting to study the initiation and propagation of cracks over time. We analyze how the creation of micro-cracks changes a pore structure over time, influencing the pore structure's total porosity. In \Cref{sec:comparison}, we employ the presented methodology to compare an actual pore space to its approximated counterpart and highlight the effect approximation of pore space has on different physical mechanisms. We show that preserving the porosity as well as the pore shape is important for the reliable modelling of corrosion-driven failure mechanisms.  Finally, in \Cref{sec:conclusion}, we conclude our study by discussing the possible applications of the proposed numerical framework, especially in modelling reaction-diffusion-driven internal cracking in porous media.

\section{Methodology}
\label{sec:method}

This section presents the numerical framework for corrosion-driven mechanisms (see \Cref{fig:conceptual}) based on the Fourier-based spectral method. Applying the Fast-Fourier Transform (FFT) to solve partial differential equations makes spectral methods more efficient than conventional methods. Since an FFT-based approach requires a pixel-based representation of the structure, its application to tomographic scans of porous media is straightforward. 

We consider a porous structure $\Omega$ in $n-$dimensional space, where $n \in [1, 2, 3]$. The solid phase is denoted as $\Omega_s$ and the pores as $\Omega_p$. The volume fraction of the pore space is denoted as $\eta = \Omega_{p}/\Omega$. The porous structure is subjected to an overall concentration gradient $\gradient c_{\mathrm{mac}}$ and an overall strain $\varepsilon_{\mathrm{mac}}$. In a multi-scale setting, $\gradient c_{\mathrm{mac}}$ and $ \tensor{\varepsilon}_{\mathrm{mac}}$ represent macroscale quantities and $\Omega$ the underlying representative volume element (RVE)~\cite{geers_multi-scale_2010}. In the next few sections, we employ FFT-based methodology to calculate the micro-scale quantities, which are concentration of diffusive species, stresses and fracture.  We choose the FFT-based Galerkin method~\cite{vondrejc_fft-based_2014} in which the gradients of concentration and displacement are the primary degrees of freedom. 

\subsection{Diffusion process}

The diffusion of ions within the pore phase $\Omega_p$ (see stage I in \Cref{fig:conceptual}) is simulated by solving the static diffusion equation over the whole domain $\Omega$ until the average concentration gradient is equal to the applied overall concentration gradient, \ie{} $\avg{ \gradient c(\vector{x}) } = \gradient c_{\mathrm{mac}}$, which is given by Fick's law as,

\begin{align}\label{eq:diffusion}
\gradient \cdot \vector{j}(\vector{x})= 0, \quad \ni \langle \gradient c(\vector{x})  \rangle= \gradient c_{\mathrm{mac}}
\end{align} 

where $\vector{j}$ is the flux defined as $\vector{j} = -\tensor{D} \cdot \gradient c(\vector{x})$ and $\tensor{D}$ is the diffusion tensor at each spatial point $\vector{x}$ defined as 

\begin{align}
  \tensor{D} = \omega D_\mathrm{solid}\tensor{I} + (1-\omega)D_{\mathrm{pore}} \tensor{I},\quad \text{where}~ \begin{cases}
    \omega =  1 &  \forall \vector{x} \in \Omega_s \\
    \omega =  0 &  \forall \vector{x} \in \Omega_p \\
    \end{cases}
\end{align}  

To solve the diffusion equation using the spectral method, the weak form of the equation is reformulated such that the unknown quantity is $\gradient c(\vector{x})$. The derivation of the weak form for \Cref{eq:diffusion} (identical to other elliptic problems such as for static equilibrium in solids) is covered in great detail in the literature \cite{zeman_finite_2017, de_geus_finite_2017}. Therefore, only the main fundamentals essential to the proposed framework are discussed here. With $\gradient c(\vector{x})$ satisfying periodic conditions, the weak form is given as

\begin{align}
\int_{\Omega} \delta \gradient c \cdot \vector{j}(\vector{x}) d\Omega = 0~,
\end{align}
where $\delta \gradient c$ is the test function. A $2^{\mathrm{nd}}$-order projection operator $\tensor{G}$ imposes the compatibility conditions (periodic and curl vanishes) on $\delta\gradient c$.  For details on the operator $\tensor{G}$, please refer to \ref{app:projection-operator}. The projection is calculated through a convolution operation between $\tensor{G}$ and an arbitrary vector $\delta\widetilde{\gradient c}$,  denoted as  $\big(\tensor{G} \ast \delta \widetilde{\gradient c}\big)(\vector{x})$ = $\int_{-\infty}^{\infty} \tensor{G}(\vector{x}):\delta\widetilde{\gradient c}(\vector{x} -\vector{y})~\mathrm{d}\vector {y}$. The weak form is now given as 

\begin{align}\label{eq:diffusion-weak-form}
\int_{\Omega} (\tensor{G} \ast\delta \widetilde{\gradient c})(\vector{x}) \cdot \vector{j}(\vector{x})~ \mathrm{d}\Omega=     \int_{\Omega} \delta \widetilde{\gradient c}(\vector{x}) \cdot (\tensor{G} \ast\vector{j} )(\vector{x}) ~\mathrm{d}\Omega= 0   
\end{align}
where the symmetry of operator $\tensor{G}$ is used. The domain is discretized into $n$ grids along each direction with discretization length $\Delta = l/n$, where $l$ represents the side length. Employing the Galerkin approach~\cite{de_geus_finite_2017, zeman_finite_2017}, the unknown continuous fields, \ie{} $\gradient c(\vector{x})$ and $\delta \widetilde{ \gradient c}(\vector{x})$, are approximated by multiplying discrete values $\gradient c(\vector{x}_k)$ and $\delta \widetilde{ \gradient c}(\vector{x}_k)$  with shape functions $\mathcal{N}_k$ defined at $n$ grid points, \ie{} 
$\gradient{c}(\vector{x}) = \sum_{k=1}^{n}\gradient{c}(\vector{x_k})\mathcal{N}(\vector{x}_k)$ and $\delta\widetilde{\gradient c}(\vector{x}) = \sum_{k=1}^{n}\delta\widetilde{\gradient c}(\vector{x_k})\mathcal{N}(\vector{x}_k)$. Upon applying the discretization, the weak form of \Cref{eq:diffusion-weak-form} can be written as:
\begin{equation}
\int_{\Omega} \underbrace{\sum_{k} \mathcal{N}(\vector{x}_k)\delta \widetilde{\gradient c}(\vector{x}_k)}_{[\delta\widetilde{\gradient c}]^{\mathrm{T}}:[\mathcal{N}]} \cdot (\tensor{G}\ast\vector{j})(\vector{x}_k)~\mathrm{d}\Omega = 0
\end{equation}
where $[\star] $ represents a column vector ( $n \times 1$ ) of quantity $\star$ evaluated at each grid points. The above equation must hold for any $\delta\widetilde{\gradient c}$, therefore, $\int_{\Omega}[\mathcal{N}]\cdot(\tensor{G}\ast\vector{j})(\vector{x}_k)\mathrm{d}\Omega=0$. A trapezoidal scheme, similar to \cite{de_geus_finite_2017} is chosen for the integration, whereby nodal points $\vector{x}_k$ serve as integration points with equal weights. This simplifies the approximated weak form, which reads:

\begin{align}\label{eq:discretized}
    \sum_{k=1}^{n}[\mathcal{N}] (\vector{x}_k) (   \tensor{G}\ast\vector{j})(\vector{x}_k) = 0 ~.
\end{align}

In the above equation, convolution is performed in Fourier space and then transformed back to the real space, which reads 

\begin{align}\label{eq:diffusion-final}
\ifft{ \fft{\tensor{G} } : \fft{ \vector{j}} } = \vector{0} ~,
\end{align}
where $\fft{f}$ represents the Fourier transform $\mathcal{F}$ of a field $f$ and $\mathcal{F}^{-1}$ inverse transform to real space. For a given overall concentration gradient $\gradient c_{\mathrm{mac}}$, the above equation is solved using an iterative solver such as a Conjugate Gradient solver. For a non-linear behaviour of $\vector{j}(\gradient c(\vector{x}))$, the solution of \Cref{eq:diffusion-final} requires Newton-Raphson iterations in addition to a Conjugate gradient solver~\cite{de_geus_finite_2017}. \Cref{alg:corrosion} explains the numerical algorithm for the solution of \Cref{eq:diffusion-final} using a Newton-Raphson iteration along with a linear iterative solver. Finally, the solution to \Cref{eq:diffusion-final} yields the concentration gradient $\gradient c(\vector{x})$ within the microstructure.  To estimate corrosion products' precipitation (discussed later in \Cref{sec:precipitation}), one requires concentration values $c(\vector{x})$ of diffusive species. Therefore, we now discuss the methodology to construct concentration $c(\vector{x})$ from the computed $\gradient c(\vector{x})$. The concentration gradient in the representative volume is expressed as 

\begin{align}\label{eq:micro-gradient}
    \gradient c(\vector{x}) = \gradient c_{\mathrm{mac}} + \gradient \phi(\vector{x})
\end{align}
where $\gradient \phi(\vector{x})$ is the periodic fluctuation of the concentration gradient due to the presence of heterogeneities in the micro-structure~\cite{noauthor_microporomechanics_2006, to_fft_2020-1}. Integrating the above equation gives the concentration of diffusive species within the microstructure, which reads:

\begin{align}\label{eq:concentration}
    c(\vector{x}) =  \gradient c_{\mathrm{mac}} \cdot\vector{x} + \phi(\vector{x}) + \beta
\end{align}
where $\beta$ is the integration constant, and $\phi(\vector{x})$ is the micro-fluctuations in the concentration of ions caused by the presence of heterogeneities. Both contributions are unknown. 
First, we determine $\phi(\vector{x})$ by solving the derivative equation of \Cref{eq:micro-gradient}:

\begin{align}\label{eq:phi-fft}
    \Delta \phi(\vector{x}) - \gradient \cdot( \gradient c(\vector{x})) = 0
\end{align}
where we used that $\gradient \cdot \left(\gradient c_\mathrm{mac} \right) = 0$, and the local concentration gradient $\gradient c(\vector {x})$ is known (from the solution of \Cref{eq:diffusion} and \Cref{eq:diffusion-final}). Then, considering $\beta$, we note from \Cref{eq:concentration}  and \Cref{eq:phi-fft} that we can choose any arbitrary value for $\beta$ without affecting the solution to \Cref{eq:diffusion-final}. We chose the value of $\beta$, so the local concentration $c(\vector{x})$ in the pores and the solid phase satisfy two conditions: (i) the average concentration in the solid phase must be zero \ie{} $ \langle c(\vector{x}) \rangle_{\mathrm{solid}} = 0$ and (ii) the average concentration in the pores must be equal to the macro-concentration of the ions divided by the porosity $\eta$, \ie{} $ \langle c(\vector{x})\rangle_{\mathrm{pore}} = c(\vector{X})/\eta$~\cite{noauthor_microporomechanics_2006}. The above two conditions apply only to a system where the solid phase has a negligible diffusion coefficient compared to the pores. On averaging \Cref{eq:concentration} over each phase and then plugging in the required average concentration values for each phase, we find two different values of $\beta$:
\begin{subequations}\label{eq:psi-two-values}
\begin{equation}
\beta_\mathrm{pore} =  c(\vector{X})/\eta - \langle \nabla c_\mathrm{mac}\cdot \vector{x} \rangle_\mathrm{pore} - \langle \phi(\vector{x})\rangle_\mathrm{pore}
\end{equation}
 \begin{equation}
\beta_\mathrm{solid} =  - \langle \nabla c_\mathrm{mac}\cdot \vector{x} \rangle_\mathrm{solid} - \langle \phi(\vector{x})\rangle_\mathrm{solid} 
\end{equation}
\end{subequations}
Substituting $\beta_\mathrm{pore}$ into \Cref{eq:concentration} yields the concentration within the  pores as
\begin{align}\label{eq:concentration-correct}
       c(\vector{x}) =  \gradient c_{\mathrm{mac}}\cdot\vector{x} + \phi(\vector{x}) +  c(\vector{X})/\eta - \langle \nabla c_\mathrm{mac}\cdot \vector{x} \rangle_\mathrm{pore} - \langle \phi(\vector{x})\rangle_\mathrm{pore}, \quad \mathrm{for~} \vector{x} \in \Omega_\mathrm{pore} ~.
\end{align}
Thus, the derived ionic concentration $c(\vector{x})$ can be employed to model various chemical reactions that will lead to the precipitation of corrosion products. The above methodology ensures that the average values of $\avg{\beta(\vector{x})}$ and $ \avg{\phi(\vector{x})}$ are 0 such that the average of local concentration $c(\vector{x})$ thus obtained is equal to the macroscopic concentration \ie{}  $\avg{c(\vector{x})} = c(\vector{X})$.


In the next section, we present the methodology to estimate the corrosion product concentration caused by the ionic concentration $c(\vector{x})$ (see stage II in \Cref{fig:conceptual}). This will later be used to compute the pressure developed due to their growth within the pores.

\subsection{Precipitation of corrosion products and pressurization of
 pores}\label{sec:precipitation}
 After acquiring the concentration of ferrous ions within the pore space, we  estimate the precipitation of corrosion products. The complex chemical reactions and phase changes in this process require thermodynamically consistent modelling for cementitious materials. In the literature, various empirical relations or thermodynamic-consistent methods, namely the Gibbs energy minimization method or the law of mass-action method,  exist for calculating precipitation. In principle, all approaches seek local chemical equilibrium between the chemical species to estimate the concentration of corrosion products. Since such approaches have been well-established, we do not discuss them here, and a reader can find a detailed description of various methodologies elsewhere~\cite{kulik_gem-selektor_2012, leal_computational_2016, furcas_solubility_2022}. For our methodology, one can choose any of these approaches to determine the precipitate concentration at each grid point; irrespective of the approach employed, the proposed methodology remains the same. For this paper, we employ a thermodynamically-consistent approach to estimate the precipitation of corrosion products. The growth of the corrosion products leads to the pressurization of pores. In this paper, we consider the pressurization of pores due to the expansion of the corrosion products.  We calculate an isotropic eigenstrain due to the expansion of corrosion products at a spatial point  $\vector{x}$, which is given as:

\begin{align}\label{eq:eigen}
\tensor{\varepsilon}_\mathrm{eig}(\vector{x}) = \dfrac{\big(V_\mathrm{ppt}(\vector{x})-V(\vector{x})\big)_{+} }{V(\vector{x})}\tensor{I}, \quad \text{where}~  \begin{cases}
    \big( \star \big )_{+} =  \star &  \text{if} ~\star > 0  \\
    \big( \star \big )_{+} =  0 &  \text{if}~ \star < 0 \\
    \end{cases}
\end{align}
where $V_\mathrm{ppt}(\vector{X})$ is the precipitate's volume, $V(\vector{x}) $ is the volume associated with the grid point $\vector{x}$ and $\big( \star  \big)_{+}$ represents Macaulay brackets. Expressing the volume of the precipitate in terms of concentration simplifies the relation,  which reads:

\begin{align}\label{eq:eigen-simplify}
\tensor{\varepsilon}_\mathrm{eig}(\vector{x}) = \big( c(\vector{x})\mathcal{M}_{\mathrm{ppt}}-1 \big)_{+} \tensor{I}
\end{align}
where $\mathcal{M}_{\mathrm{ppt}}$ is the molar volume of the precipitate. In the next section, we apply these eigenstrains within the pore space and solve for the stresses developed.

\subsection{Stress development}
Under these eigenstrains,  we solve for static mechanical equilibrium in $\Omega$ until the average strain within the microstructure is equal to the overall applied macro strain $\tensor{\varepsilon}_{\mathrm{mac}}$. The strong form equation for static mechanical equilibrium is thus given as 

\begin{align}\label{eq:solid}
\gradient \underbrace{\mathbb{C}:(\tensor{\varepsilon}(\vector{x}) -\tensor{\varepsilon}_{\mathrm{eig}}(\vector{x}) )}_{\tensor{\sigma}} = 0, \quad \ni \avg{\tensor{\varepsilon}(\vector{x})} = \tensor{\varepsilon}_{\mathrm{mac}}  ~.
\end{align}

The stiffness tensor $\mathbb{C}$ at a point $\vector{x}$ is defined as 

\begin{align}
\mathbb{C}(\vector{x}) = \omega\mathbb{C}_{\mathrm{solid}} + (1-\omega)\mathbb{C}_{\mathrm{pore}}, \quad \text{where}~ \begin{cases}
    \omega =  1 &  \forall \vector{x} \in \Omega_s \\
    \omega =  0 &  \forall \vector{x} \in \Omega_p \\
    \end{cases}
\end{align}
where for a given phase $\mathbb{C}_{i} = \lambda_{i} \tensor{I}\otimes\tensor{I} + 2 \mu_{i}\mathbb{I}$, and $\lambda_{i}$ and $\mu_{i}$ being the Lame's constants for phase $i$. The reformulation of the mechanical problem using the spectral method is similar to the diffusion problem formulation described earlier. Thus, the discretized weak form is given as:

\begin{align}\label{eq:solid-final}
\ifft{ \fft{\mathbb{G}_s}  :   \fft{\tensor{\sigma}}} = \vector{0}
\end{align}
where $\mathbb{G}_s$ represents a $4^{\mathrm{th}}$-order projection operator that imposes compatibility conditions on $\tensor{\varepsilon}(\vector{x})$ (for further details on $\mathbb{G}_s$,  please refer to ~\ref{app:projection-operator}). Since the above equation is subjected to both eigenstrains within the pores and an overall strain $\tensor{\varepsilon}_{\mathrm{app}}$, we solve the equation using a Newton-Raphson approach until the residual $\tensor{r}(\vector{x}) =  -\nabla\tensor{\sigma}( \tensor{\varepsilon}(\vector{x}) - \tensor{\varepsilon}_\mathrm{eig}(\vector{x}))$ approaches 0. Within each iteration step of the Newton-Raphson solver, \Cref{eq:solid-final} is solved using a linear iterative solver. \Cref{alg:corrosion} summarizes the numerical algorithm for solving the above equation subjected to eigenstrains within pores and an overall strain. Next, we determine the crack's initiation and propagation based on the strain state in the solid phase (see stage III in \Cref{fig:conceptual}).

\subsection{Crack initiation and propagation}

We chose a variational phase-field approach \cite{marigo_gradient_2014, Miehe2010} for modelling fracture within the solid phase. A sharp crack interface is regularized over a finite length $l_0$ where the damage within this regularized length is represented by a variable $d$ that varies from 0 (unbroken) to 1 (completely broken). For this paper, we chose the hybrid anisotropic formulation \cite{Ambati2014} for the evolution of damage variable $d$ whose strong form reads:

\begin{align} \label{eq:crack}
    -\dfrac{\mathcal{G}_c l_0}{2}\gradient \cdot\gradient d + \dfrac{\mathcal{G}_c}{2l_0} d + \mathcal{H}^{+}d = \mathcal{H}^{+} ~.
\end{align}

The term $\mathcal{H}^{+}$ is the history field that stores the maximum strain energy at a point throughout a simulation, \ie{} $\mathcal{H}^{+} = \text{max}_{ \{\tau \in [0, t] \} } \psi^{+}(\tensor{\varepsilon}^{+}(\vector{x}, \tau))$. The total elastic energy $\psi$ is decomposed to its positive $\psi^{+}$ and negative $\psi^{-}$ parts where only the strain energy associated with tension and shear, \ie{} $\psi^{+}$ (see \Cref{eq:energy}) contributes to the creation of cracks~\cite{Ambati2014,Amor2009}. The formulation thus prevents the formation of cracks in the compressed region and also, the penetration of cracks surface upon crack closure. The positive part of the strain energy density $\psi^{+}$ is given as:

\begin{align}\label{eq:energy}
\psi^{+}(\tensor{\varepsilon}) = \dfrac{1}{2}( \lambda + \mu)\big [ \text{tr}(\tensor{\varepsilon})\big]_{+}^2 + \mu(\tensor{\varepsilon}^{\text{dev}}: \tensor{\varepsilon}^{\text{dev}})
\end{align}
where $\lambda, \mu$ are Lame's coefficients, $\big[ \star \big]_{+} = \dfrac{1}{2}(\star + \| \star\|)$ and $\tensor{\varepsilon}^{\text{dev}} = \tensor{\varepsilon} - \dfrac{1}{2}\text{tr}(\tensor{\varepsilon})\tensor{I}$. 

In \Cref{eq:crack}, the parameter $l_0$ represents the regularized length scale, and $\mathcal{G}_c$ represents the fracture energy of the material. We compute the Laplacian of $d$, \ie{} $\gradient\cdot\gradient d$ in Fourier space and transform it back to the real space, hence \Cref{eq:crack} becomes
\begin{align} \label{eq:crack-fft}
    -\dfrac{\mathcal{G}_c l_0}{2}\ifft{ \mathbf{i}\vector{\xi} \cdot \mathbf{i}\vector{\xi}~ \fft{d} } + \dfrac{\mathcal{G}_c}{2l_0}d(\vector{x})  + \mathcal{H}^{+}d(\vector{x}) = \mathcal{H}^{+}  ~,
\end{align}
which is then solved using a linear iterative solver such as GMRES~\cite{noauthor_gmres_nodate}. The stiffness tensor $\mathbb{C}$ in \Cref{eq:solid} is updated to account for the reduction in stiffness due to micro-cracks within the solid. The degraded stiffness tensor at a point is given as $((1-d)^2 + \kappa)\mathbb{C}(\vector{x})$, where $\kappa$ is a small artificial residual  stiffness of the completely broken solid phase to keep \Cref{eq:solid} well-posed as $d$ approaches 1. 

\subsection{Combined methodology for corrosion-driven fracture}

\Cref{alg:corrosion} summarizes the entire methodology for the corrosion-driven fracture using an FFT-based method. We use a staggered solution scheme, which solves each mechanism in a sequential manner. We first solve the diffusion problem for given overall boundary conditions and then use the local concentrations obtained to estimate corrosion products' concentration. Similarly, the mechanical problem is solved for the obtained eigenstrains first, so strains $\tensor{\varepsilon}(\vector{x})$ are known  when the phase-field problem is solved for the initiation and propagation of cracks. The staggered scheme thus requires a small time step to minimize numerical errors. Despite this restriction, the staggered scheme allows for a straightforward implementation of the multi-physics problems.

\begin{algorithm}[H]
  \caption{FFT-based algorithm for corrosion-driven fracture problems}\label{alg:cp-algorithm}
  \begin{algorithmic}[1]  
    \State For a given overall concentration gradient $\gradient c_{\mathrm{mac}}$ and an overall strain $\tensor{\varepsilon}_{\mathrm{mac}}$
    \newline
    \newline
    \textbf{Diffusion of chemical species}
    \State $r(\vector{x}) = -\nabla\cdot\vector{j}(\gradient c_{\mathrm{mac}}),~ \gradient c(\vector{x})^{i} = \gradient c_{\mathrm{mac}}$ \Comment{$\vector{j}(\gradient c ) = \tensor{D} \cdot \gradient c$}
    \State  solve : $\ifft{ \fft{\tensor{G}} : \fft{\vector{j}(\delta \gradient c)} } = \ifft{\fft{\tensor{G}} : \fft{\vector{j}(r)} }$ \Comment{Linear iterative solver}
    \State update : $\gradient c(\vector{x})^{i+1} = \gradient c(\vector{x})^{i} + \ifft{ \fft{\delta \gradient c}}$  
    \newline
    \State solve : $\phi( \vector{\xi}) =  \gradient(\vector{\xi})   \gradient c(\vector{\xi}) /  \gradient(\vector{\xi}) .\gradient(\vector{\xi}), \quad \phi(\vector{x}) = \ifft{  \fft{\phi}} $ \Comment{Periodic fluctuations}
    \newline
    \State solve : $ c(\vector{x}) =  \gradient c_{\mathrm{mac}}\cdot\vector{x} + \phi(\vector{x}) +  c(\vector{X})/\eta - \langle \nabla c_\mathrm{mac}\cdot \vector{x} \rangle_\mathrm{pore} - \langle \phi(\vector{x})\rangle_\mathrm{pore}, \quad \mathrm{for~} \vector{x} \in \Omega_\mathrm{pore}$ \Comment{Concentration}
    \newline 
    \newline
    \textbf{Pressurization of pores}
    \State solve : precipitation at each spatial point $\vector{x}$ 
    \State compute : eigen strains $\tensor{\varepsilon}_\mathrm{eig}(\vector{x})$
    \newline
    \newline
    \textbf{Development of stresses}
        \State $\tensor{r}(\vector{x}) = -\nabla\tensor{\sigma}(\tensor{\varepsilon}_{\mathrm{mac}}),~ \tensor{\varepsilon}(\vector{x})^i = \tensor{\varepsilon}_{\mathrm{mac}}$ \Comment{$\tensor{\sigma}(\tensor{\varepsilon}) = \mathbb{C}:\tensor{\varepsilon}$}
    \While {true}
    \State  solve : $\ifft{ \fft{\mathbb{G}_s} : \fft{\tensor{\sigma}(\delta \tensor{ \varepsilon)}}  } = \ifft{\fft{\mathbb{G}_s} : \fft{\tensor{\sigma}(\tensor{r})} }$ \Comment{Linear iterative solver}
    \State update : $\tensor{\varepsilon}(\vector{x})^{i+1} = \tensor{\varepsilon}(\vector{x})^{i} + \ifft{\fft{\delta \tensor{\varepsilon}}}$  
    \If {$\| \delta \tensor{\varepsilon}(\vector{x})\| <$  tol} {break}
    \Else
    \State $\tensor{r}(\vector{x}) =  -\nabla\tensor{\sigma}( \tensor{\varepsilon}(\vector{x})^{i+1} - \tensor{\varepsilon}_\mathrm{eig}(\vector{x}))$
    \EndIf
    \EndWhile
    \newline
    \newline
    \textbf{Crack initiation and propagation}
    \State solve : $-\dfrac{\mathcal{G}_c l_0}{2}\ifft{ \mathbf{i}\vector{\xi}.\mathbf{i}\vector{\xi}~ \fft{d}} + \dfrac{\mathcal{G}_c}{2l_0}d(\vector{x})  + \mathcal{H} d(\vector{x}) = \mathcal{H} $ \Comment{Linear iterative solver}
    \State update: stiffness tensor $\mathbb{C}$  \Comment{Material degradation}
    
\end{algorithmic}
\label{alg:corrosion}
\end{algorithm}

 \subsection{Gibbs ringing artefacts}\label{sec:gibbs}
 
\begin{figure}
  \centering
\begin{tikzpicture}
  \node[inner sep=0pt] at (0, 0) {\includegraphics[trim=0cm 0cm 0cm 0cm, clip, width=\linewidth]{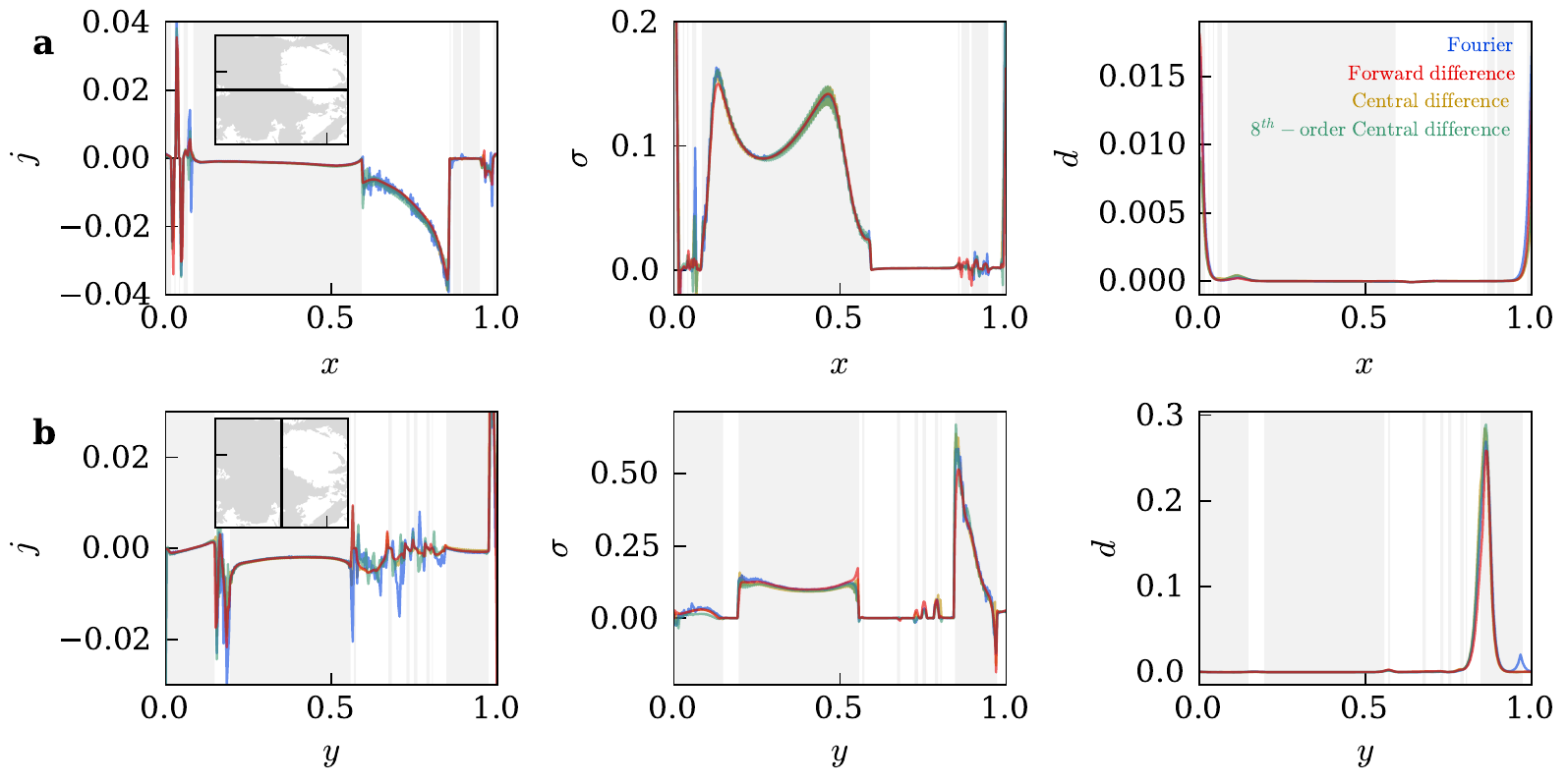}};
\end{tikzpicture}
\caption{\textbf{Effect of gradient operators on different corrosion-relevant mechanisms.} Two cross-sections along a microstructure are considered and computed with different gradient operators. The grey area represents the solid, and the white presents the pore. (a) Fluxes, stresses and damage values (from left to right) along a horizontal cross-section (marked by line in inset). (b) Fluxes, stresses and damage values  (from left to right) along a vertical cross-section (marked by line in inset). The material contrast between the pores and the solid is $10^{3}$, \ie{} $D_{pore}/D_{solid}=10^3$ and $E_{solid}/E_{pore}=10^{3}$. The results shown are for an overall concentration gradient $\gradient c_{\mathrm{mac}}=1$ and an overall strain $\tensor{\varepsilon}_{\mathrm{mac}}$ = $10^{-2}$ }
\label{fig:operator-effect}
\end{figure}

 The spectral-based method employs trigonometric basis functions, which are continuous with global support. Therefore, any contrast in local properties leads to Gibbs ringing artefacts (results show high-frequency oscillations or checkerboard patterns around the discontinuities).  In \Cref{fig:operator-effect}, we show these artefacts for different corrosion-relevant mechanisms (details of specific simulation are provided in \Cref{sec:results}) when Fourier-based gradient operator $\mathbf{i}\vector{\xi}$ is employed (represented by the blue curve in the sub-figures).  Various discrete operators with local support have been proposed in the literature to mitigate Gibbs ringing artefacts. A few of such operators are the central difference operator $ \mathbf{i}\xi_\alpha =  (\mathbf{i}\sin(\xi_\alpha\Delta)-1)/\Delta$, a higher-order central difference operator ($8^{th}$) and the forward difference operator $\mathbf{i}\xi_\alpha =  (\exp(\mathbf{i}\xi_\alpha \Delta) -1)/\Delta$. In the expressions described above, $\Delta$ represents the grid size in real space, and $\alpha$ represents the spatial direction. We employ these operators to simulate corrosion-related mechanisms and compare their effectiveness in reducing Gibbs artefacts observed in fluxes, stresses and damages. Overall, the high-frequency oscillations or Gibbs artefacts are considerably suppressed by the aforementioned discrete operators (see \Cref{fig:operator-effect}). However, we observe that the forward-difference gradient operator is the most effective among all the considered operators for mitigating Gibbs' ringing artefacts. For all mechanisms considered, it leads to smooth transitions across the discontinuities. Therefore, in the following, we will use the forward-difference gradient operator for simulations of corrosion-driven fracture. 

\section{Corrosion-driven fracture in cementitious material}\label{sec:results}

This section employs the presented numerical framework to simulate corrosion-driven fracture in a cementitious material. We consider an actual 2D scan of a cementitious sample with a porosity of $\eta$ = 0.36 obtained using focused ion beam scanning electron microscopy (see \Cref{fig:corrosion-driven-cracking}a). The sample has dimensions of $[0, l] \times [0, l]$, where $l=1$ mm, and it is discretized into $199^2$ grid points. An odd number of grid points is chosen to maintain the compatibility of the concentration gradient and deformation gradient (see \cite{zeman_finite_2017, leute_elimination_2022} for details). The free diffusion coefficient of ferrous ions within the saturated pores is taken as $D_{\mathrm{pore}}=1$ $\mathrm{mol/mm^2s}$. For a porous micro-structure, the diffusion coefficient within the solid phase should be 0. However, a negligible value of $D_{\mathrm{solid}} \approx 0$ leads to numerical difficulty in convergence, and any finite value leads to non-physical diffusive fluxes and concentrations within the solid. Therefore, we conducted a sensitivity analysis to determine an optimal value for the free diffusion coefficient in the solid phase (see \ref{sec:sensitivity}). We note that a value of $D_\mathrm{solid} = D_{\mathrm{pore}}/10^{6}$ leads to optimal results. The material parameters for the solid phase are chosen as: elastic modulus $E=10~$MPa, Poisson ratio $\nu=0.2$, critical strength $\sigma_c = 10$ MPa, and critical fracture energy $\mathcal{G}_c = 10$ J/m$^2$. The regularized length $l_0 = 0.01$ mm for the phase-field simulation is chosen based on the relation $\sigma_c=9\sqrt{E\mathcal{G}_c/6l_0}/16$~\cite{Borden2012}. Since pores are saturated with water, we consider material parameters of water $E=1$ MPa and $\nu=0.49$ for the pore space. To demonstrate the capabilities of our approach, we consider precipitation of \ce{Fe(OH)_2} only, but there are no limitations to the type of precipitation. At a constant pH of 8, as soon as the local concentration of ferrous ions exceeds the solubility limit of $10^{-3}$ mol/L, chemical reactions lead to the precipitation of \ce{Fe(OH)_2}~\cite{furcas_solubility_2022,Stefanoni2018}. Thus, the concentration of precipitated \ce{Fe(OH)_2} is computed by subtracting the solubility limit $\mathcal{S}$(pH) $= 10^{-3}$ mol/L from the total concentration of ferrous ions \ie{} $c(\vector{x})-\mathcal{S}$(pH). Meanwhile, the concentration of ferrous ions saturates to $\mathcal{S}$(pH). The material parameters for \ce{Fe(OH)_2} considered are: density $\rho=3.4$ g/cm$^3$ and molar volume $\mathcal{M} = 26$ cm$^3$/mol.

\subsection{Different corrosion-mechanisms within a micro-structure}\label{sec:fractal-single}
\begin{figure}
  \centering
\begin{tikzpicture}
\node[inner sep=0pt] at (0, 0) {\includegraphics[trim=0cm 0cm 0cm 0cm, clip, width=\linewidth]{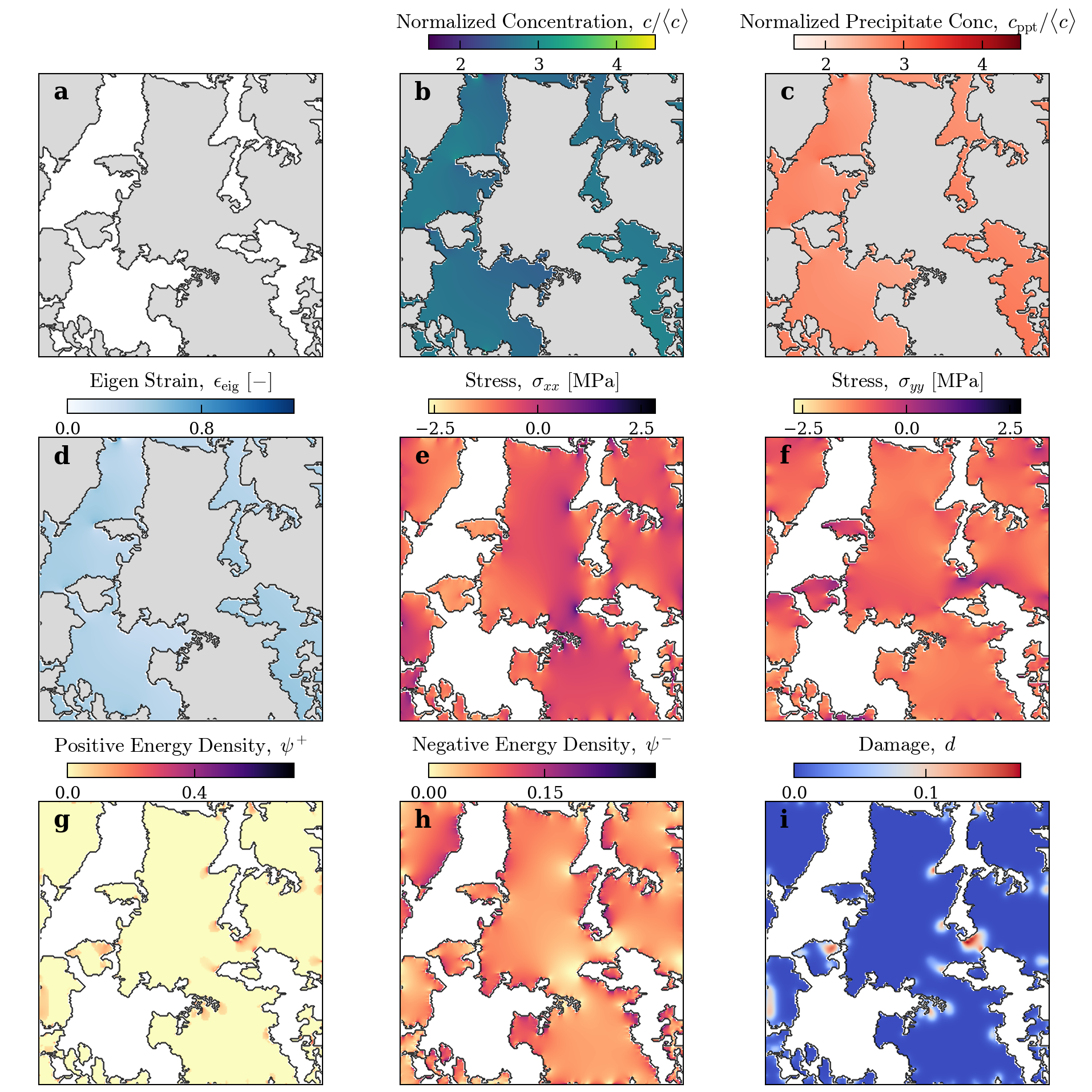}};
\end{tikzpicture}
\caption{\textbf{Corrosion-driven mechanisms in fractal pore space.} (a) The microstructure from a scan of a cementitious sample. (b) The concentration of ferrous ions (computed from \Cref{eq:concentration-correct}) within the pore space assuming an average concentration of $\avg{c} = 2\times10^{-5}~$mol/mm$^3$. Ferrous ions concentration is normalized by macro-concentration $\avg{c}$. (c) The concentration of precipitate $\ce{Fe(OH)_2}$ ($c_\mathrm{ppt}$) at a constant pH of 8 and solubility limit of $10^{-9}~$mol/mm$^3$. Concentrations are normalized by $\avg{c}$. (d) Eigenstrains $\tensor{\varepsilon}_{eig}$ developed due to the expansion of $\ce{Fe(OH)_2}$ within the pore space, computed from~\Cref{eq:eigen-simplify}. (e,f) Stress components $\sigma_{xx}$ and $\sigma_{yy}$, respectively, computed from \Cref{eq:solid} and \Cref{eq:solid-final}.  (g,h) The positive and negative components of the elastic energy, respectively. contributing to the crack initiation and propagation is computed from \Cref{eq:energy}. (i) Crack initiation and propagation within solid phase represented by phase-field variable $d$.}
\label{fig:corrosion-driven-cracking}
\end{figure}

We simulate the physical process within the microstructure under static conditions, \ie{} at a certain instance in time. We assume that the sample is subjected to an overall concentration gradient of $\gradient c_{\mathrm{mac}} = 10^{-6}~$mol/mm and an overall strain of $\tensor{\varepsilon}_{\mathrm{mac}} = \tensor{0}$ at a constant pH of 8. Furthermore, we arbitrarily chose a macro-concentration $\avg{c}$ value (superimposed to the gradient) to compute the micro-concentrations within the pore space (see \Cref{eq:concentration-correct}). The values of $\gradient c_\mathrm{mac}$ and $\avg{c}$ are chosen to ensure that the local concentration of ferrous ions $c(\vector{x})$ reach the solubility limit $\mathcal{S}$(pH) $= 10^{-3}$ mol/L (see \Cref{fig:corrosion-driven-cracking}b) and $\ce{Fe(OH)_2}$ precipitates within pores (see \Cref{fig:corrosion-driven-cracking}c), which leads to eigenstrains induced by the volume expansion of $\ce{Fe(OH)_2}$ (see \Cref{fig:corrosion-driven-cracking}d). Due to the fractal nature of the microstructure, we observe the concentration of stresses along the pore boundaries (see \Cref{fig:corrosion-driven-cracking}e and \Cref{fig:corrosion-driven-cracking}f).
 We observe the initiation of several micro-cracks along the pore boundaries (see \Cref{fig:corrosion-driven-cracking}i), which result from stress concentration along the pore interface. 

We note that this 2D simulation with 199$\times$199 grid points, performed on a single-core machine, takes approx 5 seconds for the diffusion problem, 15 seconds for the elasticity problem, and 9 seconds for solving the phase-field equation, totalling 29 seconds. The majority of the simulation time $\approx$ 90$\%$ is spent in constructing the two projection operators $\mathcal{G}$ and $\mathbb{G}$). 

\begin{figure}
  \centering
\begin{tikzpicture}
\node[inner sep=0pt] at (0, 0) {\includegraphics[trim=0cm 0cm 0cm 0cm, clip, width=0.85\linewidth]{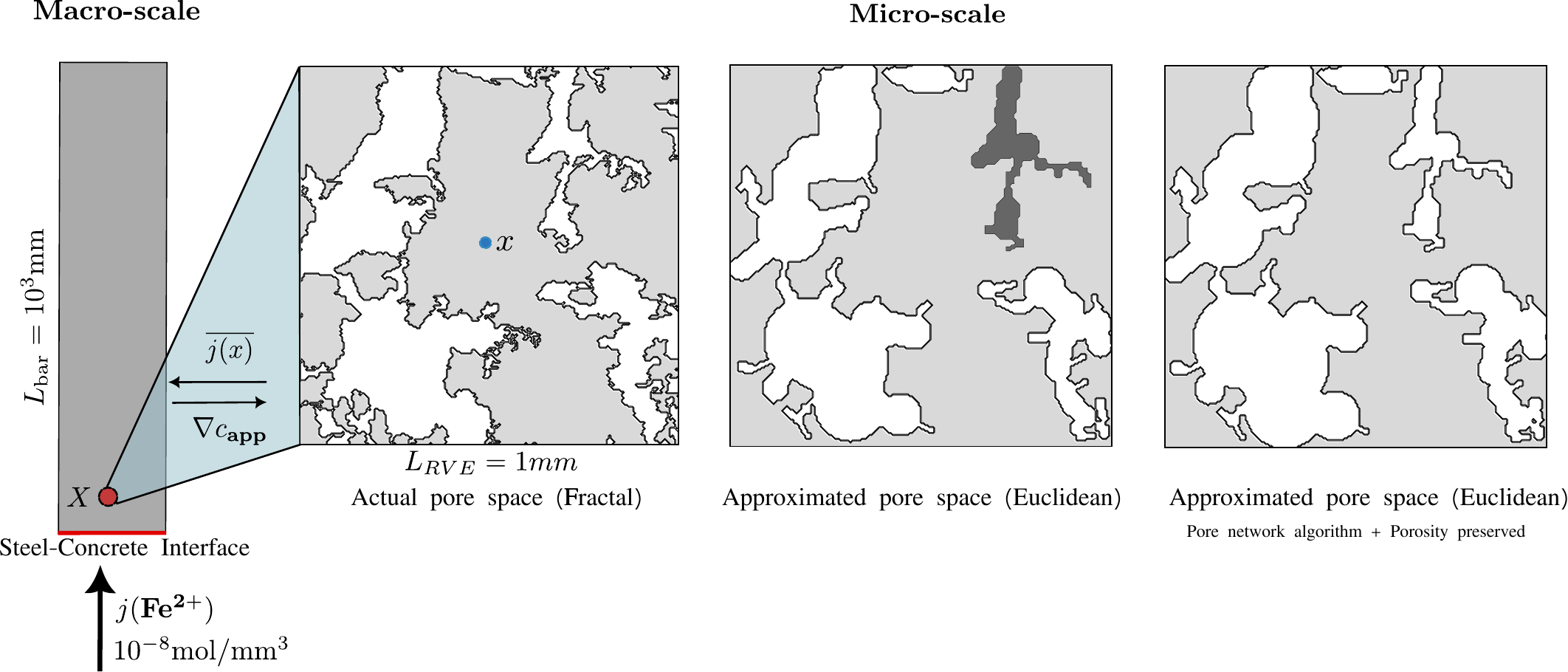}};
\end{tikzpicture}

\caption{\textbf{Multiscale approach for diffusion process:} Schematic of employed multiscale approach for \Cref{sec:evolution}. (left) Each material point $\vector{X}$ at the macroscale is coupled to a representative volume. The lower side of the bar is subjected to a constant flux of ferrous ions. The macroscale concentration gradients are passed to the coupled RVE to compute the diffusive flux (from \Cref{eq:diffusion-final}). The local diffusive flux $\vector{j}(\vector{x})$ is then averaged over an RVE and is passed to the macroscale to perform the diffusion at the macroscale. The macroscale concentration $c(\vector{X})$ thus computed is passed to the RVE to compute local concentrations of precipitates within the pores. (middle) Approximated pore space generated using the pore-network algorithm that does not preserve porosity. A few pore spaces, shown in dark grey, are isolated from the boundaries (including the top boundary) and thus do not contribute to the diffusion process. (right) Pore space is generated from the modified pore-network algorithm that preserves total porosity. }

\label{fig:multiscale-schematic}
\end{figure}

\subsection{Evolution of micro-cracks within a micro-structure}\label{sec:evolution}

Next, we apply the proposed methodology to study how stresses and, subsequently, the micro-cracks develop within a micro-structure over time. To simulate the temporal evolution of the corrosion-driven mechanisms, we employ a multi-scale setting (see \Cref{fig:multiscale-schematic}). To this end, we consider a bar composed of cementitious material whose lower surface is subjected to a constant flux of ferrous ions (see \Cref{fig:multiscale-schematic}-left). The lower surface is representative of a Steel-Concrete Interface (SCI) where steel corrosion happens in cementitious materials. The fluxes are zero at other boundaries. Each material point of the bar, represented as $\vector{X}$, is coupled to a representative volume element (RVE), in this case, the fractal porous media considered previously. Here, we use the same microstructure for every point, but there are no technical limitations to generate different microstructures for each point. The diffusion coefficients ($D_{\mathrm{pore}}, D_{\mathrm{solid}}$) for an RVE are considered as described previously. The diffusion process is simulated using a multi-scale approach where macroscopic concentration gradients are applied to the RVEs, in which the effective diffusive flux \ie{} $\avg{\vector{j}(\vector{x})}$ is computed and then upscaled to the macro scale, where it is used to compute the macroscopic concentration of ferrous ions in the bar. The multi-scale approach is only used for the diffusion process; all other processes are simulated only locally within the RVEs. At the macro-scale, we solve \Cref{eq:diffusion} using a finite difference scheme. 
For a constant pH system, as considered in this study, the precipitation occurs as soon as the ferrous ion concentration reaches the solubility limit $\mathcal{S}(\mathrm{pH})$. Finally, the stresses within the solid phase are computed assuming a zero overall strain \ie{} $\tensor{\varepsilon}_{\mathrm{mac}}$ = 0 at each time step. The material parameters ($E, \nu, \sigma_c, \mathcal{G}_c$, $l_0$) are considered as described previously. 

\begin{figure}
  \centering
\begin{tikzpicture}
\node[inner sep=0pt] at (0, 0) {\includegraphics[trim=0cm 0cm 0cm 0cm, clip, width=\linewidth]{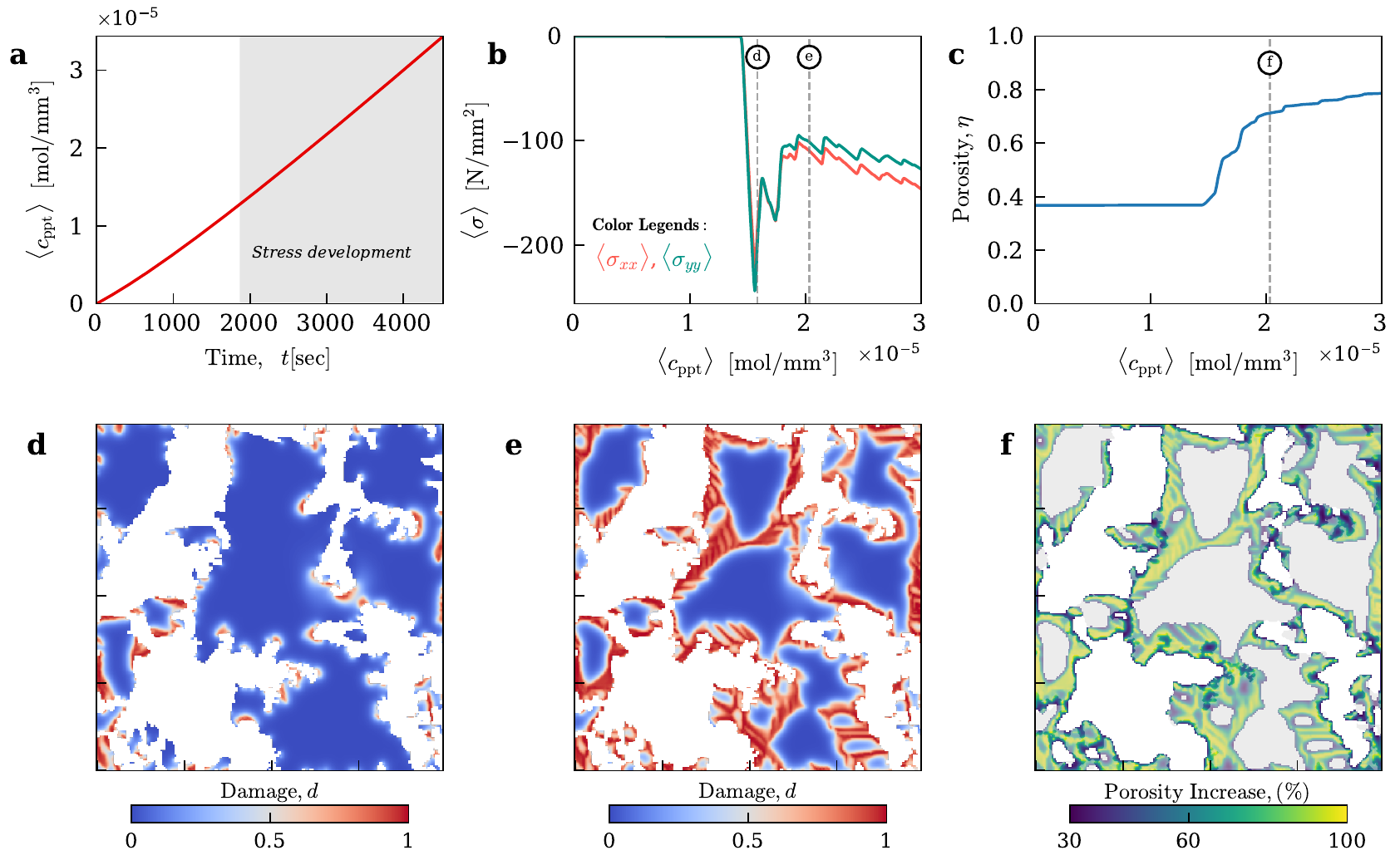}};

\end{tikzpicture}
\caption{\textbf{Evolution of damage, average stress and total porosity.} Different corrosion-driven mechanisms are shown for the RVE located at $5~$mm from the steel-concrete interface. (a) Evolution of average concentration of corrosion product over time. (b) Development of average stresses $\avg{\sigma}_{xx}$ (red) and $\avg{\sigma}_{yy}$ (green). (c) Increase in total porosity of the RVE due to internal cracking. (d-e) Evolution of internal cracks represented by the damage $d$ within the solid phase of the RVE at two different time instances indicated by grey lines in (b). (f) State of porosity within the solid phase of the RVE at a time instance indicated by grey line in (c). }
\label{fig:multi-scale}
\end{figure}

We analyze the results for the RVE located at a distance of $5~$mm from the steel-concrete interface (see \Cref{fig:multiscale-schematic}). Unlike the corrosion products, the stresses within the solid phase of the RVE form after a delay (see \Cref{fig:multi-scale}a and \Cref{fig:multi-scale}b). The stresses develop after 1.5 $\times 10^{-5}~$mol/mm$^3$ of corrosion products have precipitated. The delay is caused because locally, a precipitate $V_\mathrm{ppt}(\vector{x})$ must expand to the associated cell volume $V(\vector{x})$ to exert pressure (see \Cref{eq:eigen}). Therefore, a precipitate must reach a threshold concentration locally to exert pressure. The initial drop in the average stresses is due to micro-crack initiation (see \Cref{fig:multi-scale}d-e) and the relaxation caused due to it. The initiation and propagation of micro-cracks also affect other mechanisms. For example, internal cracks increase the porosity of a solid phase (see \Cref{fig:multi-scale}c) and thus facilitate the diffusion of chemical species further into the solid phase. To this end, we use the damage variable from the phase field to estimate the increase in the porosity of a fractured region, $d=0$ being non-porous and $d=1$ being completely porous (see \Cref{fig:multi-scale}f). The total porosity of the RVE increases as the micro-cracks initiate and propagate within the solid phase of the RVE (shown by the blue curve in \Cref{fig:multi-scale}c). Although in this study, we do not consider the effect of porosity change on other mechanisms (for example, diffusion), the micro-structure changes must be coupled to the other physical mechanism. Since the quantities, such as the concentration of ferrous ions and damage values per grid point, are stored in a standard vector or matrix format (no special data structures are required), the coupling of the process through data exchange is straightforward. Furthermore, a regularized description of fracture also allows for coupling the fracture process with the ingress of chemical species due to increased porosity, which makes the study of the interplay among mechanisms feasible \cite{Wu2016}.

\begin{figure}
  \centering
\begin{tikzpicture}
\node[inner sep=0pt] at (0, 0) {\includegraphics[trim=0cm 0cm 0cm 0cm, clip, width=0.6\linewidth]{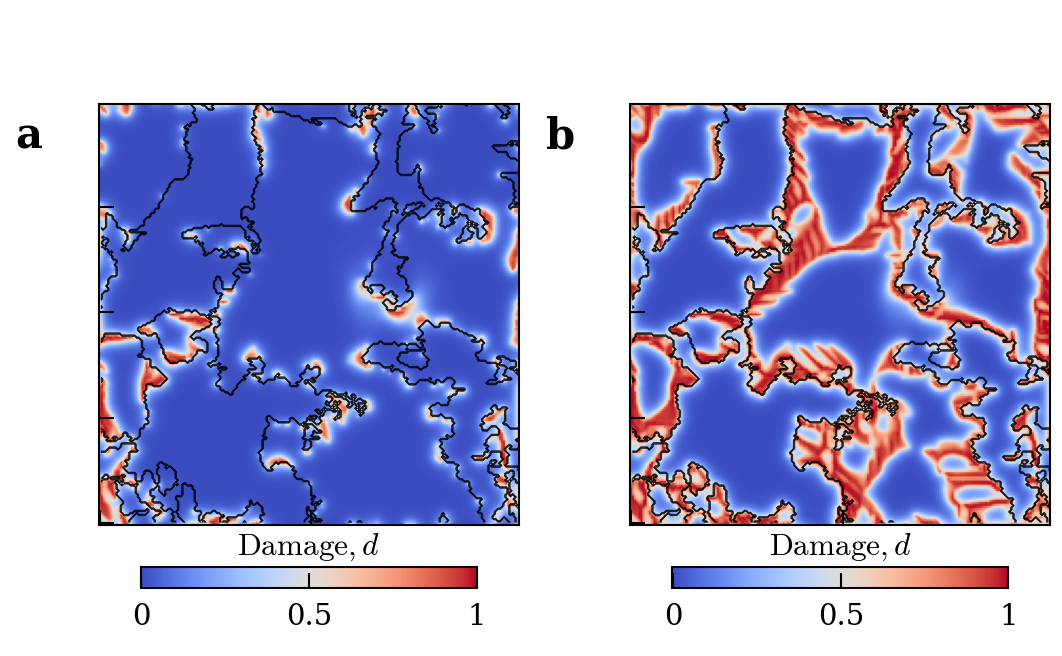}};
\end{tikzpicture}
\caption{\textbf{Numerical artefacts of phase-field formulation.} Damage variable is shown over the entire domain of the 
microstructure, \textit{i.e.} in the solid and pores, for the fractal pore structure at times corresponding to (a) \Cref{fig:corrosion-driven-cracking}d  and (b) \Cref{fig:corrosion-driven-cracking}e.}
\label{fig:fracture-in-pores}
\end{figure}

Although the phase-field formulation employed in the numerical framework is robust in handling crack initiation and propagation and subsequently, coupling to other physical processes, it also leads to some non-physical observations. For example, a closer look at \Cref{fig:operator-effect} and \Cref{fig:fracture-in-pores} shows that cracks also develop within the pores. This is mainly because of the diffusive representation of a sharp crack over a length $2l_0$. When a crack originates within the solid phase along the interface, due to its finite width of $2l_0$, a part of it also develops within pores. However, such cracks within the pores never further develop or lead to crack propagation within the pore phase (as seen in \Cref{fig:fracture-in-pores}).  Therefore, the development of cracks within the pores is a numerical artefact of the phase-field formulation employed, which can be reduced by taking a smaller value of the length scale parameter $l_0$.   

Finally, we note that the above simulation took approximately 4 hours when performed on a single-core machine.    

\section{Comparison with a Euclidean pore space}\label{sec:comparison}

The main advantage of an FFT-based methodology is that it preserves the actual microstructure. As previously mentioned, conventional numerical frameworks approximate the fractal pore space to reduce computational complexities. This considerably approximates the various physical processes that lead to corrosion-driven fracture. 

\begin{figure}
  \centering
\begin{tikzpicture}
\node[inner sep=0pt] at (0, 0) {\includegraphics[trim=0cm 0cm 0cm 0cm, clip, width=\linewidth]{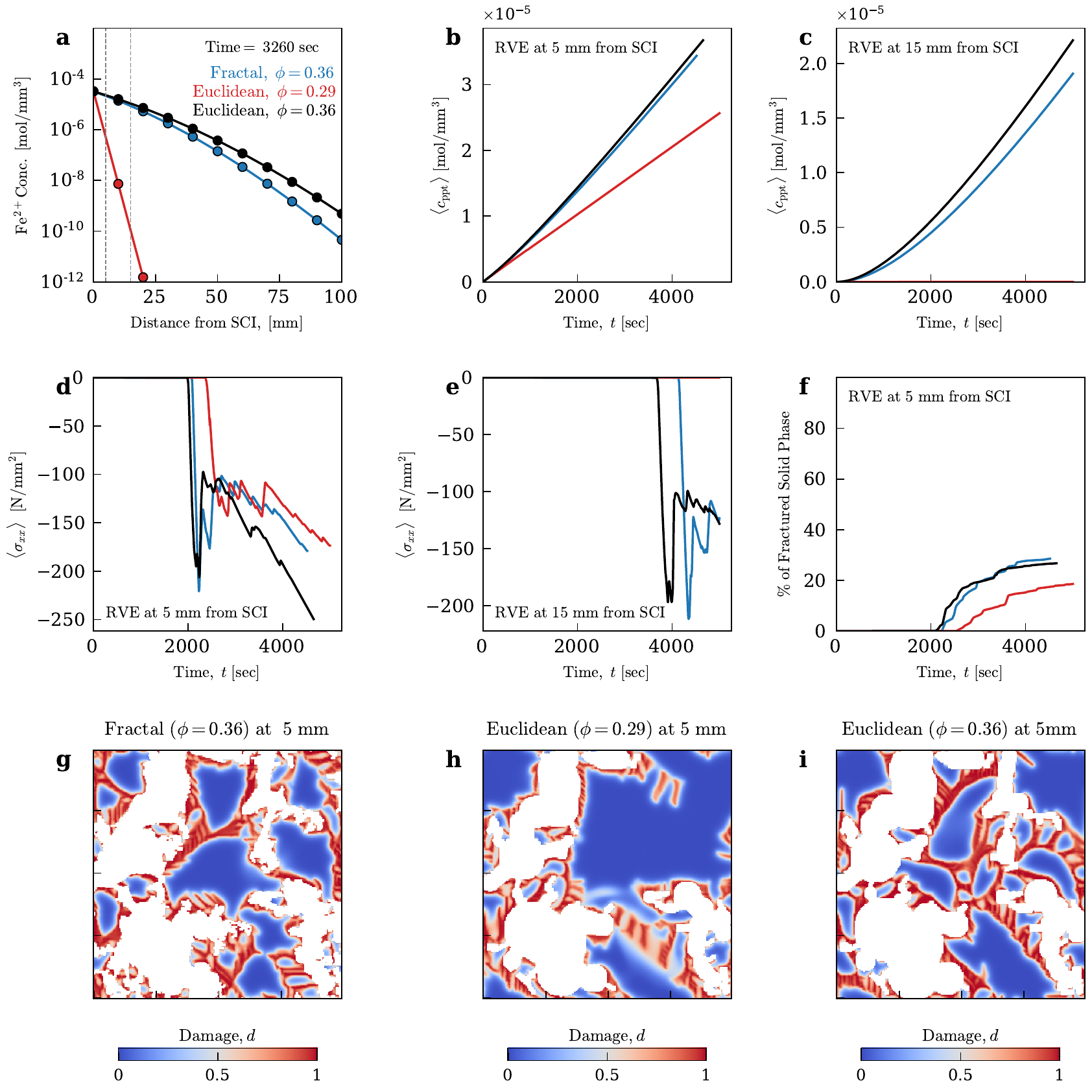}};
\end{tikzpicture}
\caption{\textbf{Comparison with Euclidean pore space.} (a) Ferrous ions concentration profile along the length of the bar after 3260 sec. The colours indicated the three different RVEs considered: actual pore space (blue), approximated pore space generated from the pore-network algorithm (red) and approximate pore space with the same porosity as actual pore space (black). (b-c) Amount of corrosion product precipitated within RVEs. The two RVEs considered are 5 mm and 15 mm from the steel-concrete interface. The location of the two RVEs is indicated with grey dotted lines in (a). The value shown for the concentration is averaged over an RVE. (d-e) Development of stresses $\avg{\sigma}_{xx}$ at the same two RVEs. The stresses value shown are averages of stresses within the solid. (f) Percentage of damaged solids within the RVE located at 5 mm. A material point within an RVE is considered completely damaged once the damage variable value exceeds $0.9$. (g-i) State of micro-cracks  within the three types of RVE considered. All the states shown are at time = 3260 sec.}
\label{fig:multi-scale-diffusion}
\end{figure}

To demonstrate the effect of pore space approximation on corrosion-driven mechanisms, we simplify the 2D fractal pore space considered  in \Cref{sec:results} and repeat the multi-scale simulation with the same conditions as the actual pore space. We use the pore-network approach~\cite{gostick_versatile_2017}, employed primarily for studying flow or diffusion in porous media, to simplify the complex pore space using spheres and cylinders (see  \Cref{fig:multiscale-schematic}). The approximation of the pore space results in lower total porosity ($\eta$=$0.29$) compared to the fractal pore space ($\eta=0.36$), and also the isolation of certain pore space from the boundaries (see dark gray in \Cref{fig:multiscale-schematic}-middle). Since only open pores contribute to the diffusion process, we set the diffusion coefficient of isolated pores to that of the solid. The effect of reduced pore space can be seen in the diffusion of ferrous ions along the bar and the precipitation of corrosion within the pores (see \Cref{fig:multi-scale-diffusion}a and \Cref{fig:multi-scale-diffusion}b). As expected, an approximated pore space, with reduced porosity, allows less diffusion of ferrous ions compared to the fractal pore space and consequently has a slower precipitation rate than the actual pore space (see \Cref{fig:multi-scale-diffusion}b for the RVE located at 5 mm from SCI). For an RVE located further away from the SCI, the concentration of diffused ferrous never reaches the solubility limit. Hence, precipitation never occurs (see \Cref{fig:multi-scale-diffusion}c for the RVE located at 15 mm). Compared to a fractal pore space, the stresses develop much later in an approximated pore space due to the slow precipitate rate (see \Cref{fig:multi-scale-diffusion}d). The delay is significant as we go further away from the SCI (see \Cref{fig:multi-scale-diffusion}e where stresses have not developed for an approximated pore space). Furthermore, the maximum value of the average stress $\bar{\sigma}_{xx}$ in the solid before the crack initiation is also different. The maximum stress increases by a factor two from the fractal pore space to the approximated pore space. Additionally, the way the micro-cracks initiate and propagate within the solid phase is entirely different in the two cases (see \Cref{fig:multi-scale-diffusion}g-h). 

\begin{figure}
  \centering
\begin{tikzpicture}
\node[inner sep=0pt] at (0, 0) {\includegraphics[trim=0cm 0cm 0cm 0cm, clip, width=\linewidth]{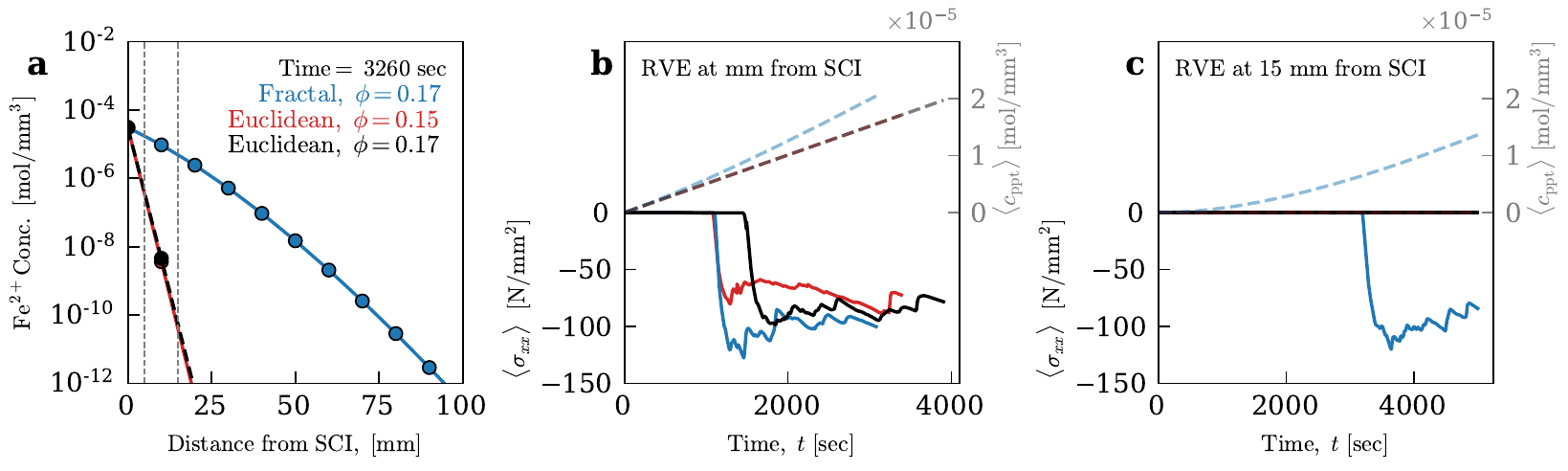}};

\end{tikzpicture}
\caption{\textbf{Comparison with Euclidean pore space at low porosity.} (a) Ferrous ions concentration profile along the length of the bar after 3260 sec. The colours indicated the three different types of RVE considered: actual pore space (blue), approximated pore space generated from the pore-network algorithm (red) and approximate pores space with the same porosity as actual pore space (black). (b-c) Evolution of stresses within the solids of an RVE. The results in (b) and (c) are for RVEs located 5 mm and 15 mm from the steel-concrete interface. The stresses shown are $\avg{\sigma}_{xx}$, averaged within the solid of an RVE. The dotted lines in the figures indicate the precipitate's concentration and evolution over time within an RVE. The values shown are average values.}
\label{fig:multi-scale-diffusion-low-porosity}
\end{figure}

In the previous example, we employed the pore-network algorithm, which inscribes spheres and cylinders within a pore space for approximation. As a result, the total porosity is not preserved, and the differences observed in the previous section are mainly due to the variation in porosity. To analyze the influence of pore shape, we again approximate the fractal pore space but preserve the porosity. To this end, we increase the radii of the inscribed spheres until the approximated pore space has the same porosity as that of fractal space \ie{}  $\eta=0.36$. \Cref{fig:multiscale-schematic}(right) shows the approximated pore space with the same porosity. Preserving the porosity leads to a similar concentration profile of diffused ions along the bar. Consequently, a similar precipitation rate within the RVEs located at $Y={5}$ mm and $15$ mm (see \Cref{fig:multi-scale-diffusion}b and \Cref{fig:multi-scale-diffusion}c). Although the maximum stress value ($\avg{\sigma}_{xx} \approx $-$200$ MPa for RVEs at $5$ mm and $15$ mm) and the start of failure ($t \approx 2000$ sec for RVE at $5$ mm and $t\approx$ 4000 sec for RVE at $15$ mm) are approximately the same, the evolution of stresses post-cracking is significantly different. The post-cracking stresses in approximated pore space ($\eta=0.36$) are much higher compared to actual pore space (see \Cref{fig:multi-scale-diffusion}d and  \Cref{fig:multi-scale-diffusion}e). The undamaged solid phases (indicated by blue colour in \Cref{fig:multi-scale-diffusion}g-i), which are under compression as precipitates grow within the pores, contribute to the post-cracking stresses. The variation in micro-cracks propagation within the two spaces (see \Cref{fig:multi-scale-diffusion}g-i) significantly affects how the undamaged spaces are created and how the stresses develop locally within such undamaged areas. Given that the same percentage of solid phase has been damaged ($\approx$ 30$\%$  see \Cref{fig:multi-scale-diffusion}f) in the two spaces, the different post-cracking stresses highlight the necessity for accurate resolution of local stresses and micro-cracks. The effect of pore shape approximation is much more distinct for pore spaces with low porosity. \Cref{fig:multi-scale-diffusion-low-porosity} compares mechanisms for a pore space with a porosity of $\eta=0.17$. Unlike for high porous spaces, preserving the total porosity does not result in the same diffusion profile, possibly due to the isolation of pores from the boundaries. As expected, the differences in the diffusion of ions lead to different precipitation rates and the development of stresses within RVEs. Especially for the RVE located far from the steel-concrete interface at $15$ mm (see \Cref{fig:multi-scale-diffusion-low-porosity}c).

In the previous two sections, we employ the proposed methodology to show the effect that an approximation of the pore space and shape has on corrosion-driven mechanisms.  A detailed study that characterizes the exact differences in various mechanisms that arise due to the approximation of pore space and shape is beyond the scope of this paper and will be considered for future work. 

\clearpage

\clearpage

\clearpage

\clearpage

\section{Discussion}\label{sec:discussion}

Although cementitious materials have been around for centuries, little to no attention has been given to the influence of materials' micro-structure on the corrosion-driven propagation phase of failure.  The focus has been mainly on macroscopic properties, such as overall porosity or electrical resistivity. As we showed in our analysis, having the same porosity but different microstructure (\Cref{fig:corrosion-driven-cracking} and \Cref{fig:multi-scale-diffusion-low-porosity}) leads to different stress evolution in the microstructure. This also directly affects how crack initiate and propagate at the micrometre scale, which in a later stage will impact other corrosion-driven mechanisms. Another measure often employed to include micro-scale features in predicting corrosion-related mechanisms is the pore-size distribution obtained experimentally. However, measurements obtained from common techniques assumes that pores are either cylindrical or spherical. As we showed, approximating a complex micro-structure with smoother features can lead to a different behaviour of corrosion-driven mechanisms.
Furthermore, measures such as overall porosity or the pore size distribution lack information regarding the geometry and spatial arrangement of pores or, conversely, solid phases, which is critical for understanding local stress state and crack initiation (see \Cref{fig:multi-scale}). Therefore, numerical analyses employing these measures provide limited insight into the underlying corrosion-driven mechanisms because they lack a fundamental consideration of the effect of microstructure. However, we consider such fundamental consideration important, especially in the context of the significant endeavors towards fabricating novel cementitious materials by chemically altering micro-structure~\cite{zunino_microstructural_2022}. The presently proposed framework will thus contribute to a better understanding of the microstructure-related features critical to corrosion-driven damage. Principally, determining the desired values of such critical features/measures could increase the serviceability lifetime of concrete structures in corrosive environments. We believe the proposed framework will thus help tailor or design sustainable cementitious materials.

To identify the features of a microstructure that are critical to corrosion-driven damage, a detailed statistical analysis of three-dimensional RVEs is necessary. This remains difficult as it requires a complete analysis where both diffusion as well as stresses are solved in a multiscale setting. Moreover, since the features of a microstructure will evolve (due to the deposition of precipitates and formation of crack channels as shown in \Cref{fig:corrosion-driven-cracking}), both processes must be coupled. Even though the FFT framework, as proposed here, is computationally efficient (as can be seen by the computational cost reported in \Cref{sec:fractal-single} and \Cref{sec:evolution}),  its efficiency suffers when the property contrast between phases is significantly high. This asks for computationally efficient solvers or algorithms, compared to the ones employed here, capable of handling large differences in phase properties. The optimization of the algorithm with respect to computational efficiency is beyond the scope of this paper and will be considered for future work.

Lastly, the proposed numerical framework could be improved by validating it with experimental data, although specifically designed experiments are needed to provide the required validation data. In the present study, we assume certain conditions or parameters largely due to a lack of experimental values. For example, the corrosion rate or diffusive flux value employed in \Cref{sec:evolution} is assumed to be constant along the entire steel-concrete interface. However, under conditions representative for engineering structures, the corrosion rate is variable as it depends on time-variable exposure conditions, and ultimately the electrochemically active steel surface \ie{} steel surface in contact with water. Recent studies~\cite{stefanoni_electrochemistry_2018} estimate this variability that could be incorporated  into the numerical framework for better predicting corrosion-driven mechanisms. Similarly, in the present framework, we assumed two distinct values of $\beta$ in \Cref{eq:psi-two-values} (\ie{} $\beta_\mathrm{pore}$ and $\beta_\mathrm{solid}$) to compute local concentrations within the pores. This numerical trick ensures that the local concentration values satisfy certain macro-scale conditions. However, the physical significance behind this $\beta$ parameters remains unknown, which should be verified by comparison with experimental observations. Furthermore, we also assume that the entire pore space of an RVE is saturated with water. A better estimation of the  condensation of micro-pores would be helpful in further improving the proposed numerical framework. We believe these assumptions or limitations of the proposed numerical framework can be surmounted by actively verifying the model with experimental results and designing experiments to elucidate details of the microstructure of cementitious materials.

\section{Conclusion}\label{sec:conclusion}

This paper presents a numerical framework for fracture within fractal porous media where the interplay among different physical mechanisms drives the fracture. The proposed FFT-based spectral integral methodology is computationally efficient and easy to implement for fractal pore spaces, and its extension to a multi-scale approach is straightforward. We highlight the capability and the robustness of the spectral-integral method to resolve different physical problems, \textit{e.g.}, diffusion and mechanical, within complex micro-structure, such as fractal pore spaces in cementitious materials. We show the significance of actual pore space by drawing a comparison with an approximated pore space, as usually done in the literature. Our comparisons show that the approximation of pore space may severely underestimate various physical mechanisms (e.g. reactive transport) and fracture initiation and propagation. The present methodology thus opens a path for a better understanding of the effect of complex pore spaces, an aspect often neglected or crudely approximated, on corrosion-driven mechanisms in cementitious materials. Although this paper mainly focuses on corrosion-driven fracture in concrete, the presented methodology is general. It can be easily extended to other multi-physics-driven fractures in complex porous media. 

\section{Acknowledgements}
The authors would like to thank Thilo Schmid (Durability of Engineering Materials, ETH Zurich) for providing the FIB-SEM scan of the cementitious samples used in this study.

\section{Data Availability}
The code~\cite{noauthor_cmbm-public_nodate} for the simulation is written in Python and is an extension of the code provided in \cite{de_geus_finite_2017}. The simulation data generated in this study have been deposited in the ETH Research Collection database under accession code \href{https://doi.org/10.3929/ethz-b-000593923}{ethz-b-000593923}.

\newpage
\appendix

\section{Projection operator} \label{app:projection-operator}
 The main objective of the projection operator ($\tensor{G}$ or $\mathbb{G}$) in the FFT Galerkin approach is to project an arbitrary tensor into a compatible one.  For a diffusion problem, the compatible tensor is the concentration gradient ($1^\mathrm{st}$-order) and for the elasticity problem,  the compatible tensor is the strain tensor ($2^\mathrm{st}$-order). The convolution operations $\tensor{G}\ast\gradient c$ and $\mathbb{G}\ast\tensor{\varepsilon}$ in the Fourier space can be expressed as:
 \begin{subequations}
 \begin{equation}
\fft{A_i} = \fft{g_{ij}}\fft{\nabla c_j},
\end{equation}
\begin{equation}
~~ \fft{A_{ij}} = \delta_{im}\fft{g_{jl}}\fft{\varepsilon_{ml}}
 \end{equation}
 \end{subequations}
 
 where $\fft{A_i}$ and $ \fft{A_{ij}}$ are the respective compatible tensors in Fourier space.  For a diffusion problem, the concentration gradient $\vector{\nabla}c(\vector{x})$ is only a gradient of a scalar field \ie{} the concentration of diffusive species $c(\vector{x})$. Similarly for an elasticity problem, a row vector of strain tensor is only the gradient of a component of the displacement field $\vector{u}(\vector{x})$. An arbitrary field vector $\vector{f}(\vector{x})$  is only the gradient of a scalar field $g(\vector{x})$ if its curl vanishes, \ie{} $\nabla \times \vector{f}(\vector{x}) = 0$. Furthermore, $\vector{f}(\vector{x})$ must be periodic. These two conditions thus form the compatibility conditions. Since the periodicity of $\vector{f}(\vector{x})$ is inherently satisfied by the Fourier transform, the projection operator $\fft{g_{ij}}$ is formulated such that the curl of $\vector{f}(\vector{x})$ vanishes. Leute et al~\cite{leute_elimination_2022} mathematically prove that $\fft{g_{ij}}  = \mathbf{i}\xi_{i}\cdot(\mathbf{i}\vector{\xi})^{-1}_{j} $ projects an arbitrary field into a compatible one in a least-square sense, \ie{} the residual vector $\mathcal{R} = \gradient{g}(\vector{x}) - \vector{f}(\vector{x})$ is minimized. In the expression for the projection operator, $(\mathbf{i}\vector{\xi})^{-1}$ is the inverse of the Fourier representation of the gradient given as 
\begin{align}\label{eq:g-42}
(\mathbf{i}\vector{\xi})^{-1} = \dfrac{\mathbf{i}\vector{\xi}^{\star}}{\mathbf{i}\vector{\xi}\cdot \mathbf{i}\vector{\xi}^{\star} } ~,
\end{align}
where $\mathbf{i}\vector{\xi}^{\star}$ is the conjugate of $\mathbf{i}\vector{\xi}$. The above relation applies only for $\mathbf{i}\vector{\xi}\cdot\mathbf{i}\vector{\xi}^{\star}   \neq 0$, otherwise $0$.

\section{Sensitivity analysis}\label{sec:sensitivity}

\begin{figure}[H]
  \centering
\begin{tikzpicture}
\node[inner sep=0pt] at (0, 0) {\includegraphics[trim=0cm 0cm 0cm 0cm, clip, width=0.65\linewidth]{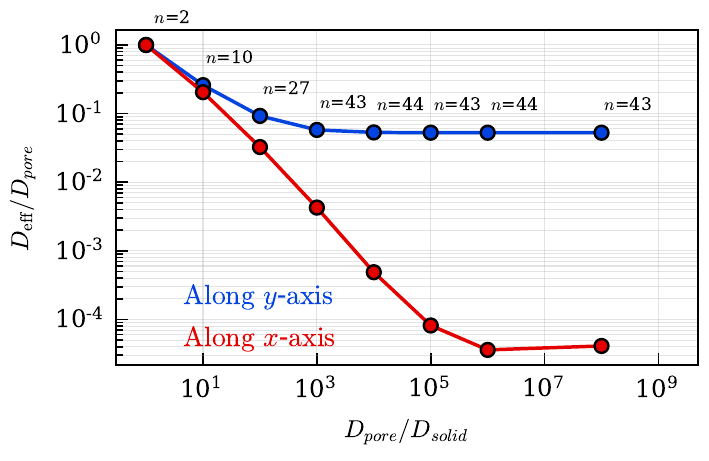}};
\end{tikzpicture}
\caption{\textbf{Sensitivity analysis:} Variation in the microstructure's effective diffusion coefficient based on the solid phase's diffusion coefficient value. The effective diffusion coefficient converges to a value of $~10^{-5}D_\mathrm{pore}$ for $D_\mathrm{solid} < 10^{-6}D_\mathrm{pore}$. Therefore, we chose a value of $10^{-6}D_\mathrm{pore}$ to represent a non-porous solid phase in \Cref{sec:results}. }
\label{fig:optimal}
\end{figure}

\newpage

\bibliography{\myreferences}

\begin{thebibliography}{10}

\bibitem{angst_challenges_2018}
U.~M. Angst, ``Challenges and opportunities in corrosion of steel in
  concrete,'' {\em Materials and Structures}, vol.~51, p.~4, Feb. 2018.

\bibitem{noauthor_corrosion_2003}
{\em Corrosion of {Steel} in {Concrete}}.
\newblock John Wiley \& Sons, Ltd, 1~ed., 2003.
\newblock \_eprint: https://onlinelibrary.wiley.com/doi/pdf/10.1002/3527603379.

\bibitem{ulm_is_2003}
F.-J. Ulm, ``Is concrete a poromechanics material? - {A} multiscale
  investigation of poroelastic properties,'' {\em Materials and Structures},
  vol.~37, pp.~43--58, Nov. 2003.

\bibitem{ioannidou_mesoscale_2016}
K.~Ioannidou, K.~J. Krakowiak, M.~Bauchy, C.~G. Hoover, E.~Masoero, S.~Yip,
  F.-J. Ulm, P.~Levitz, R.~J.-M. Pellenq, and E.~Del~Gado, ``Mesoscale texture
  of cement hydrates,'' {\em Proceedings of the National Academy of Sciences},
  vol.~113, pp.~2029--2034, Feb. 2016.

\bibitem{zhu_fractal_2019}
J.~Zhu, R.~Zhang, Y.~Zhang, and F.~He, ``The fractal characteristics of pore
  size distribution in cement-based materials and its effect on gas
  permeability,'' {\em Scientific Reports}, vol.~9, p.~17191, Nov. 2019.
\newblock Bandiera\_abtest: a Cc\_license\_type: cc\_by Cg\_type: Nature
  Research Journals Number: 1 Primary\_atype: Research Publisher: Nature
  Publishing Group Subject\_term: Civil engineering;Composites
  Subject\_term\_id: civil-engineering;composites.

\bibitem{gao_fractal_2014}
Y.~Gao, J.~Jiang, G.~De~Schutter, G.~Ye, and W.~Sun, ``Fractal and multifractal
  analysis on pore structure in cement paste,'' {\em Construction and Building
  Materials}, vol.~69, pp.~253--261, Oct. 2014.

\bibitem{pia_geometrical_2013}
G.~Pia and U.~Sanna, ``A geometrical fractal model for the porosity and thermal
  conductivity of insulating concrete,'' {\em Construction and Building
  Materials}, vol.~44, pp.~551--556, July 2013.

\bibitem{zeng_surface_2013}
Q.~Zeng, M.~Luo, X.~Pang, L.~Li, and K.~Li, ``Surface fractal dimension: {An}
  indicator to characterize the microstructure of cement-based porous
  materials,'' {\em Applied Surface Science}, vol.~282, pp.~302--307, Oct.
  2013.

\bibitem{yang_fractal_2017}
X.~Yang, F.~Wang, X.~Yang, and Q.~Zhou, ``Fractal dimension in concrete and
  implementation for meso-simulation,'' {\em Construction and Building
  Materials}, vol.~143, pp.~464--472, July 2017.

\bibitem{jilesen_three-dimensional_2012}
J.~Jilesen, J.~Kuo, and F.-S. Lien, ``Three-dimensional midpoint displacement
  algorithm for the generation of fractal porous media,'' {\em Computers \&
  Geosciences}, vol.~46, pp.~164--173, Sept. 2012.

\bibitem{zhang_determination_1995}
B.~Zhang and S.~Li, ``Determination of the {Surface} {Fractal} {Dimension} for
  {Porous} {Media} by {Mercury} {Porosimetry},'' {\em Industrial \& Engineering
  Chemistry Research}, vol.~34, pp.~1383--1386, Apr. 1995.
\newblock Publisher: American Chemical Society.

\bibitem{winslow_fractal_1995}
D.~Winslow, J.~M. Bukowski, and J.~F. Young, ``The fractal arrangement of
  hydrated cement paste,'' {\em Cement and Concrete Research}, vol.~25,
  pp.~147--156, Jan. 1995.

\bibitem{okabe_pore_2005}
H.~Okabe and M.~J. Blunt, ``Pore space reconstruction using multiple-point
  statistics,'' {\em Journal of Petroleum Science and Engineering}, vol.~46,
  pp.~121--137, Feb. 2005.

\bibitem{wang_study_2009}
Y.~Wang, T.~Zhang, J.~Liu, and J.~Zhang, ``The {Study} of {Porous} {Media}
  {Reconstruction} {Using} a {2D} {Micro}-{CT} {Image} and {MPS},'' in {\em
  2009 {International} {Conference} on {Computational} {Intelligence} and
  {Software} {Engineering}}, pp.~1--5, Dec. 2009.

\bibitem{noauthor_microporomechanics_2006}
{\em Microporomechanics}.
\newblock John Wiley \& Sons, Ltd, 1~ed., 2006.
\newblock \_eprint: https://onlinelibrary.wiley.com/doi/pdf/10.1002/0470032006.

\bibitem{geers_multi-scale_2010}
M.~G.~D. Geers, V.~G. Kouznetsova, and W.~A.~M. Brekelmans, ``Multi-scale
  computational homogenization: {Trends} and challenges,'' {\em Journal of
  Computational and Applied Mathematics}, vol.~234, pp.~2175--2182, Aug. 2010.

\bibitem{liu_multi-scale_2020}
C.~Liu, Z.~Liu, and Y.~Zhang, ``A multi-scale framework for modelling effective
  gas diffusivity in dry cement paste: {Combined} effects of surface, {Knudsen}
  and molecular diffusion,'' {\em Cement and Concrete Research}, vol.~131,
  p.~106035, May 2020.

\bibitem{maekawa_multi-scale_2014}
K.~Maekawa, {\em Multi-{Scale} {Modeling} of {Structural} {Concrete}}.
\newblock London: CRC Press, Apr. 2014.

\bibitem{moulinec_numerical_1998}
H.~Moulinec and P.~Suquet, ``A numerical method for computing the overall
  response of nonlinear composites with complex microstructure,'' {\em Computer
  Methods in Applied Mechanics and Engineering}, vol.~157, pp.~69--94, Apr.
  1998.

\bibitem{de_geus_finite_2017}
T.~W.~J. de~Geus, J.~Vondřejc, J.~Zeman, R.~H.~J. Peerlings, and M.~G.~D.
  Geers, ``Finite strain {FFT}-based non-linear solvers made simple,'' {\em
  Computer Methods in Applied Mechanics and Engineering}, vol.~318,
  pp.~412--430, May 2017.

\bibitem{leute_elimination_2022}
R.~J. Leute, M.~Ladecký, A.~Falsafi, I.~Jödicke, I.~Pultarová, J.~Zeman,
  T.~Junge, and L.~Pastewka, ``Elimination of ringing artifacts by
  finite-element projection in {FFT}-based homogenization,'' {\em Journal of
  Computational Physics}, vol.~453, p.~110931, Mar. 2022.
\newblock arXiv: 2105.03297.

\bibitem{Bonnet2017}
M.~Bonnet, {\em Boundary {Integral} {Equation} {Methods} for {Elastic} and
  {Plastic} {Problems}}.
\newblock 2017.
\newblock Publication Title: Encyclopedia of Computational Mechanics Second
  Edition.

\bibitem{vondrejc_fft-based_2014}
J.~Vondřejc, J.~Zeman, and I.~Marek, ``An {FFT}-based {Galerkin} method for
  homogenization of periodic media,'' {\em Computers \& Mathematics with
  Applications}, vol.~68, pp.~156--173, Aug. 2014.

\bibitem{zeman_finite_2017}
J.~Zeman, T.~W.~J. de~Geus, J.~Vondřejc, R.~H.~J. Peerlings, and M.~G.~D.
  Geers, ``A finite element perspective on nonlinear {FFT}-based
  micromechanical simulations,'' {\em International Journal for Numerical
  Methods in Engineering}, vol.~111, no.~10, pp.~903--926, 2017.
\newblock \_eprint: https://onlinelibrary.wiley.com/doi/pdf/10.1002/nme.5481.

\bibitem{to_fft_2020-1}
Q.-D. To and G.~Bonnet, ``{FFT} based numerical homogenization method for
  porous conductive materials,'' {\em Computer Methods in Applied Mechanics and
  Engineering}, vol.~368, p.~113160, Aug. 2020.

\bibitem{kulik_gem-selektor_2012}
D.~A. Kulik, T.~Wagner, S.~V. Dmytrieva, G.~Kosakowski, F.~F. Hingerl, K.~V.
  Chudnenko, and U.~R. Berner, ``{GEM}-{Selektor} geochemical modeling package:
  revised algorithm and {GEMS3K} numerical kernel for coupled simulation
  codes,'' {\em Computational Geosciences}, Aug. 2012.

\bibitem{leal_computational_2016}
A.~M.~M. Leal, D.~A. Kulik, and G.~Kosakowski, ``Computational methods for
  reactive transport modeling: {A} {Gibbs} energy minimization approach for
  multiphase equilibrium calculations,'' {\em Advances in Water Resources},
  vol.~88, pp.~231--240, Feb. 2016.

\bibitem{furcas_solubility_2022}
F.~E. Furcas, B.~Lothenbach, O.~B. Isgor, S.~Mundra, Z.~Zhang, and U.~M. Angst,
  ``Solubility and speciation of iron in cementitious systems,'' {\em Cement
  and Concrete Research}, vol.~151, p.~106620, Jan. 2022.

\bibitem{marigo_gradient_2014}
J.-J. Marigo, ``Gradient damage models: construction and fundamental
  properties,'' 2014.

\bibitem{Miehe2010}
C.~Miehe, F.~Welschinger, and M.~Hofacker, ``Thermodynamically consistent
  phase-field models of fracture: {Variational} principles and multi-field {FE}
  implementations,'' {\em International Journal for Numerical Methods in
  Engineering}, vol.~83, pp.~1273--1311, Sept. 2010.

\bibitem{Ambati2014}
M.~Ambati, T.~Gerasimov, and L.~De~Lorenzis, ``A review on phase-field models
  of brittle fracture and a new fast hybrid formulation,'' {\em Computational
  Mechanics}, vol.~55, no.~2, pp.~383--405, 2014.
\newblock ISBN: 0178-7675.

\bibitem{Amor2009}
H.~Amor, J.-J. Marigo, and C.~Maurini, ``Regularized formulation of the
  variational brittle fracture with unilateral contact: {Numerical}
  experiments,'' {\em Journal of the Mechanics and Physics of Solids}, vol.~57,
  2009.

\bibitem{noauthor_gmres_nodate}
``{GMRES}: {A} {Generalized} {Minimal} {Residual} {Algorithm} for {Solving}
  {Nonsymmetric} {Linear} {Systems}.''

\bibitem{Borden2012}
M.~J. Borden, C.~V. Verhoosel, M.~A. Scott, T.~J.~R. Hughes, and C.~M. Landis,
  ``A phase-field description of dynamic brittle fracture,'' {\em Computer
  Methods in Applied Mechanics and Engineering}, vol.~217-220, pp.~77--95,
  2012.
\newblock Publisher: Elsevier B.V. ISBN: 0045-7825.

\bibitem{Stefanoni2018}
M.~Stefanoni, Z.~Zhang, U.~Angst, and B.~Elsener, ``The kinetic competition
  between transport and oxidation of ferrous ions governs precipitation of
  corrosion products in carbonated concrete,'' {\em RILEM Technical Letters},
  vol.~3, pp.~8--16, 2018.

\bibitem{Wu2016}
T.~Wu and L.~De~Lorenzis, ``A phase-field approach to fracture coupled with
  diffusion,'' {\em Computer Methods in Applied Mechanics and Engineering},
  vol.~312, pp.~196--223, Dec. 2016.
\newblock Publisher: Elsevier B.V.

\bibitem{gostick_versatile_2017}
J.~T. Gostick, ``Versatile and efficient pore network extraction method using
  marker-based watershed segmentation,'' {\em Physical Review E}, vol.~96,
  p.~023307, Aug. 2017.
\newblock Publisher: American Physical Society.

\bibitem{zunino_microstructural_2022}
F.~Zunino and K.~Scrivener, ``Microstructural developments of limestone
  calcined clay cement ({LC3}) pastes after long-term (3 years) hydration,''
  {\em Cement and Concrete Research}, vol.~153, p.~106693, Mar. 2022.

\bibitem{stefanoni_electrochemistry_2018}
M.~Stefanoni, U.~M. Angst, and B.~Elsener, ``Electrochemistry and capillary
  condensation theory reveal the mechanism of corrosion in dense porous
  media,'' {\em Scientific Reports}, vol.~8, p.~7407, May 2018.
\newblock Number: 1 Publisher: Nature Publishing Group.

\bibitem{noauthor_cmbm-public_nodate}
``{CMBM}-public / {Papers} {Supplementary} {Information} / 2023 / {An}
  {FFT}-based framework for predicting corrosion-driven damage in fractal
  porous media · {GitLab}.''

\end{thebibliography}

 \end{document}